\documentclass[twocolumn]{aastex701}

\newcommand{\rbs}[0]{SN\,2025rbs\,\,}
\newcommand{\NaIDnospace}[0]{Na\,I\,D}
\newcommand{\NaID}[0]{Na\,I\,D\,\,}
\newcommand{\NaIDII}[0]{Na\,I\,D$_2$\,\,}
\newcommand{\NaIDI}[0]{Na\,I\,D$_1$\,\,}
\newcommand{\KI}[0]{K\,I\,\,}
\usepackage{hyperref}
\usepackage{CJK}
\begin{document}

\title{No Evidence for Nearby Circumstellar Material in the Type Ia Supernova 2025rbs}

\correspondingauthor{Aravind Pazhayath Ravi}
\email{apazhayathravi@ucdavis.edu}

\newcommand{\UA}{\affiliation{Steward Observatory, University of Arizona, 933 North Cherry Avenue, Tucson, AZ 85721-0065, USA}}
\newcommand{\GeminiNorth}{\affiliation{Gemini Observatory/NSF's NOIRLab, 670 N. A'ohoku Place, Hilo, HI 96720, USA;}}
\newcommand{\Catalyst}{\altaffiliation{LSSTC Catalyst Fellow}}
\newcommand{\Hubble}{\altaffiliation{NASA Hubble Fellow}}
\newcommand{\Monash}{\affiliation{School of Physics and Astronomy, Monash University, Clayton, Australia}}
\newcommand{\OzGrav}{\affiliation{OzGrav: The ARC Center of Excellence for Gravitational Wave Discovery, Australia}}
\newcommand{\UCSD}{\affiliation{Department of Astronomy \& Astrophysics, University of California, San Diego, 9500 Gilman Drive, MC 0424, La Jolla, CA 92093-0424, USA}}
\newcommand{\UCD}{\affiliation{Department of Physics and Astronomy, University of California, Davis, 1 Shields Avenue, Davis, CA 95616-5270, USA}}
\newcommand{\LCO}{\affiliation{Las Cumbres Observatory, 6740 Cortona Drive, Suite 102, Goleta, CA 93117-5575, USA}}
\newcommand{\UCSB}{\affiliation{Department of Physics, University of California, Santa Barbara, CA 93106-9530, USA}}
\newcommand{\UT}{\affiliation{Department of Astronomy, The University of Texas at Austin, 2515 Speedway, Stop C1400, Austin, TX 78712, USA}}
\newcommand{\Keck}{\affiliation{W.~M.~Keck Observatory, 65-1120 M\=amalahoa Highway, Kamuela, HI 96743-8431, USA}}
\newcommand{\CfA}{\affiliation{Center for Astrophysics \textbar{} Harvard \& Smithsonian, 60 Garden Street, Cambridge, MA 02138-1516, USA}}
\newcommand{\UNC}{\affiliation{Department of Physics and Astronomy, University of North Carolina, 120 East Cameron Avenue, Chapel Hill, NC 27599, USA}}
\newcommand{\CIERA}{\affiliation{Center for Interdisciplinary Exploration and Research in Astrophysics (CIERA), 1800 Sherman Ave., Evanston, IL 60201, USA}}
\newcommand{\NU}{\affiliation{Department of Physics and Astronomy, Northwestern University, 2145 Sheridan Road, Evanston, IL 60208, USA}}
\newcommand{\USask}{\affiliation{Department of Physics and Engineering Physics, University of Saskatchewan, 116 Science Place, Saskatoon, SK S7N 5E2, Canada}}
\newcommand{\UvA}{\affiliation{Department of Astronomy, University of Virginia, 530 McCormick Rd, Charlottesville, VA 22904, USA}}
\newcommand{\SU}{\affiliation{The Oskar Klein Centre, Department of Astronomy, Stockholm University, AlbaNova, SE-10691 Stockholm, Sweden}}
\newcommand{\Rutgers}{\affiliation{Department of Physics and Astronomy, Rutgers, The State University of New Jersey, 136 Frelinghuysen Rd, Piscataway, NJ 08854-8019, USA}}
\newcommand{\Tsinghua}{\affiliation{Physics Department, Tsinghua University, Beijing 100084, China}}
\newcommand{\Berkeley}{\affiliation{Department of Astronomy, University of California, Berkeley, CA 94720-3411, USA}}
\newcommand{\UCSC}{\affiliation{Department of Astronomy and Astrophysics, University of California, Santa Cruz, CA 95064, USA}}
\newcommand{\UF}{\affiliation{Department of Astronomy, University of Florida, 211 Bryant Space Science Center, Gainesville, FL 32611-2055, USA}}
\newcommand{\KITP}{\affiliation{Kavli Institute for Theoretical Physics, University of California, Santa Barbara, CA 93106, USA}}
\newcommand{\STScI}{\affiliation{Space Telescope Science Institute, 3700 San Martin Drive, Baltimore, MD 21218, USA}}
\newcommand{\Xinjiang}{\affiliation{Xinjiang Astronomical Observatory, Chinese Academy of Sciences, Urumqi, Xinjiang, 830011, China}}
\newcommand{\NRAO}{\affiliation{National Radio Astronomy Observatory, 520.0Edgemont Rd, Charlottesville VA 22903, USA}}
\newcommand{\NotreDame}{\affiliation{Department of Physics, University of Notre Dame, Notre Dame, IN 46556, USA}}
\newcommand{\Caltech}{\affiliation{Department of Astronomy, California Institute of Technology, Pasadena, CA 91125, USA}}
\newcommand{\NOIRLab}{\affiliation{NSF's NOIRLab, 950 N. Cherry Avenue, Tucson, AZ 85719, USA}}
\newcommand{\IPAC}{\affiliation{Caltech/IPAC, Mailcode 100-22, Pasadena, CA 91125, USA}}
\newcommand{\NESI}{\affiliation{NASA Exoplanet Science Institute / Caltech-IPAC}}

\begin{CJK*}{UTF8}{gbsn}

\author[0000-0002-7352-7845]{Aravind P. Ravi} \UCD \email{apazhayathravi@ucdavis.edu}

\author[orcid=0000-0002-0832-2974, gname=Griffin, sname=Hosseinzadeh]{Griffin Hosseinzadeh}
\UCSD \email{ghosseinzadeh@ucsd.edu}

\author[0000-0001-8818-0795]{Stefano Valenti} \UCD \email{valenti@ucdavis.edu}

\author[0000-0001-8738-6011]{Saurabh~W.~Jha} \Rutgers \email{saurabh@physics.rutgers.edu}

\author[orcid=0000-0003-0123-0062, gname=Jennifer, sname=Andrews]{Jennifer Andrews}
\GeminiNorth \email{Jennifer.Andrews@noirlab.edu}

\author[orcid=0000-0003-4102-380X, gname=David, sname= Sand]{David J. Sand}
\UA \email{dsand@arizona.edu}

\author[0000-0003-3504-5316]{Benjamin J. Fulton} \NESI \email{}

\author[0000-0002-9123-0068]{William D. Vacca}
\NOIRLab \email{}

\author[orcid=0000-0002-9154-3136, gname=Melissa, sname=Graham]{Melissa L. Graham}
\UA \email{mlg3k@uw.edu}

\author[0000-0003-3460-0103]{Alexei V. Filippenko}
\Berkeley \email{}

\author[0000-0003-4537-3575]{Noah Franz}
\UA \email{}

\author[orcid=0000-0002-0744-0047, gname=Jeniveve, sname=Pearson]{Jeniveve Pearson}
\UA \email{jenivevepearson@arizona.edu}

\author[0000-0002-1895-6639]{Moira Andrews}
\LCO \UCSB \email{moira_andrews@ucsb.edu }

\author[orcid=0000-0002-4924-444X, gname= Azalee, sname=Bostroem]{K. Azalee Bostroem} 
\IPAC \email{bostroem@arizona.edu}


\author[orcid=0000-0003-0528-202X, gname=Collin, sname=Christy]{Collin Christy}\UA \email{collinchristy@arizona.edu}

\author[0000-0002-7937-6371]{Yize Dong (董一泽)} 
\CfA 
\email{yize.dong@cfa.harvard.edu} %


\author[0000-0001-6395-6702]{Sebastian Gomez}
\UT \email{}

\author[orcid=0000-0002-1125-9187]{Daichi Hiramatsu} \UF \email{}

\author[orcid=0000-0003-2744-4755, gname=Emily, sname=Hoang]{Emily Hoang}
\UCD \email{emthoang@ucdavis.edu}

\author[0000-0003-4253-656X]{D.\ Andrew Howell}
\LCO \UCSB \email{}

\author[orcid=0000-0002-9454-1742, gname=Brian, sname=Hsu]{Brian Hsu}
\UA \email{bhsu@arizona.edu}

\author[orcid=0000-0003-0549-3281, gname=Daryl, sname=Janzen]{Daryl Janzen}
\USask \email{daryl.janzen@usask.ca}

\author[orcid=0000-0003-3108-1328, gname=Lindsey, sname=Kwok]{Lindsey A. Kwok} \Hubble
\CIERA  \email{lindsey.kwok@northwestern.edu}

\author[orcid=0000-0001-9589-3793, gname=Michael, sname=Lundquist]{Michael ~J. Lundquist}
\Keck \email{mlundquist@keck.hawaii.edu}

\author[0000-0001-5807-7893]{Curtis McCully}
\LCO \email{}

\author[orcid=0009-0008-9693-4348, gname=Darshana, sname=Mehta]{Darshana Mehta}
\UCD \email{ddmehta@ucdavis.edu}


\author[orcid=0000-0002-7015-3446, gname=Nicol\'as, sname=Meza-Retamal]{Nicol\'as Meza-Retamal}
\UCD \email{nemezare@ucdavis.edu}

\author[orcid=0000-0002-4022-1874, gname=Manisha, sname=Shrestha]{Manisha Shrestha}
\Monash \OzGrav \email{manisha.shrestha@monash.edu}

\author[orcid=0000-0001-8073-8731, gname=Bhagya, sname=Subrayan]{Bhagya Subrayan}
\UA \email{bsubrayan@arizona.edu}

\author[0000-0002-3725-3058]{Lauren Weiss}
\NotreDame \email{}

\author[0000-0002-2636-6508]{WeiKang Zheng} 
\Berkeley \email{}


\begin{abstract}
We present a high-resolution spectral time series of the Type Ia supernova (SN) 2025rbs discovered in the nearby galaxy NGC 7331. The  Automated Planet Finder (APF) at Lick Observatory and the MAROON-X/IGRINS-2 at Gemini North were used to obtain echelle spectra between $-$5 and 15 days with respect to the epoch of maximum light. Several unsaturated \NaID absorption components along the line of sight are identified, but there is no evidence of time variance in any of them. We measure the equivalent width of the observed diffuse interstellar band around 5780 \AA\ and constrain the extinction along the line of sight to SN\,2025rbs as $A_V = 0.64\,\pm\,0.32$ mag, corresponding to a moderate reddening of $E(B-V) = 0.21\,\pm\,0.10$ mag (assuming $R_\mathrm{V}$ = 3.1). The observed Ca\,II H\&K interstellar absorption roughly traces \NaID in velocity space, suggesting a common origin. Quantitative comparisons between the column densities of Na and Ca gas in these host clouds ($N_\mathrm{\NaID}$\,/\,$N_\mathrm{Ca\,II}$ of order unity) argue against their origin in the Galactic halo gas and instead support absorption due to the interstellar gas of NGC 7331. Time invariance of all the observed absorption features suggests a lack of nearby circumstellar material ($\lesssim$\,10$^{16}$ cm) around the progenitor system of SN\,2025rbs. This supports a progenitor scenario for SN 2025rbs with minimal ambient circumstellar gas, consistent with a double-degenerate CO white dwarf binary system.


\end{abstract}





\section{Introduction} \label{sec:1}

Type Ia supernovae (SNe~Ia) are thought to be thermonuclear explosions of white dwarfs (WDs) in close binary systems. SNe~Ia as ``standardizable candles" are a cornerstone of the cosmological distance-ladder formulation and act as direct probes of binary stellar evolution, yet there is much debate on the nature of the binary companion and the explosion mechanisms leading to the thermonuclear runaway \citep[see][and references therein]{Maoz14, Taubenberger17, Jha19, Liu23, Ruiter25}. Among the two broad binary progenitor systems, in the single-degenerate (SD) scenario, the WD accretes matter from a main-sequence or giant star until ignition near the Chandrasekhar mass \citep{Whelan73}, whereas the double-degenerate (DD) scenario involves mass accretion from another tidally disrupted WD \citep{Iben84, Webbink84}, or even merger of the two WDs \citep{Katz12, Kushnir13}. 

In the SD scenario, while some of the mass from the companion is accreted onto the WD, the remainder can enshroud the binary system as circumstellar matter (CSM). Even within a DD system, prior to disruption, mass transfer from the secondary to the primary WD can form an accretion disk owing to its sufficiently large angular momentum and may produce continuous outflow in the form of disk winds \citep{Zenati19b}. This outflow can expand outward to form CSM \citep{Raskin_Kasen13}, a fraction of which might accumulate around the primary WD, forming a low-mass envelope \citep{Shen12, Schwab16}. The amount, composition, and velocity of this material is dependent on the exact progenitor scenario. From an observer's perspective, this outflowing material (connected to the progenitor system) would appear blueshifted and may be close enough ($\sim$10$^{16}$--10$^{17}$ cm) to induce an evolution in the observed absorption features \citep[e.g., \NaIDnospace;][]{Patat07}. A spectroscopic time-series at a few km s$^{-1}$ velocity resolution is necessary to track these ephemeral CSM signatures. 

The first identified example of temporal evolution in high-resolution spectral features was SN Ia 2006X, where a line complex of blueshifted \NaID absorption evolved between $-$2, +14, and +61 days relative to maximum brightness \citep{Patat07}. They noted evolution in the highest velocity components, indicating a possible correlation with photoionization of CSM most recently released and closest to the SN. Two additional cases of time-variable Na (and therefore the presence of CSM) in SNe\,Ia were observed for SN\,1999cl \citep{Blondin09} and SN\,2007le \citep{Simon09}, with high- and low-extinctions along line of sight. In the nearby peculiar SN Ia PTF 11kx, a series of high-resolution spectra showed clear evolution in the blueshifted lines of Ca, Fe, He, and H, indicative of a likely SN Ia progenitor star exploding into multiple shells of nearby CSM within a symbiotic system \citep{Dilday12}. SN\,2013gh was noted to have a varying \NaID component consistent with photoionization \citep{Ferretti16}. In the nearby SN\,2014J a varying \KI line (while \NaID stayed constant) was observed and attributed to photoionization \citep{Graham15}, although \cite{Maeda16} argue that this gas is unlikely to have an origin in CSM. More systematic sample searches for multi-epoch high-resolution spectra did not reveal additional examples of temporal variance \citep{Sternberg14}, although \cite{Ferretti16} argue that almost all existing time series miss the phases where photoionization of circumstellar gas is expected. 

Other sample studies of SNe~Ia at high resolution \citep[e.g.,][]{Sternberg11, Maguire13} have shown a statistical preponderance of blueshifted features. A substantial fraction of SNe\,Ia with preferentially blue-shifted absorptions were noted to express \NaID much stronger than what would be expected from their photometric colors \citep{Phillips13}. The unusual strength and preferential blueshifts are frequently tied to the CSM shed by the progenitor systems \citep{Sternberg11}, motivating several high-resolution campaigns searching for photoionization-recombination variability expected if the outflowing circumstellar gas lies close the explosion \citep[e.g.,][]{Sternberg14, Graham15, Ferretti17b}. Although, neither of these features alone can conclusively prove the presence of CSM if the absorption strengths do not vary with time as the material gets photoionized. While a few showed temporal variation as described before, several SNe\,Ia do not, thus constraining their absorbing gas to interstellar distances by its lack of variability and photoionization exclusion limits \citep[e.g.,][]{Ferretti17a}. 

Bright and spectroscopically well-classified SNe Ia act as background light sources against which the slow-moving foreground gas and dust along our line-of-sight can be studied through narrow absorption features superposed on the broder ejecta features \citep[e.g.,][]{Sollerman05, Phillips13, Gonzalez-Gaitan24}. This is especially crucial to pin down accurate color corrections considering the role of SNe Ia as ``standardizable candles" in the cosmological distance ladder formulation \citep[e.g.,][]{Riess98}. Among these, the absorption strengths of interstellar \NaID \citep[e.g.,][]{Munari_Zwitter97, Poznanski12} and the diffuse interstellar bands (DIBs) in the Milky Way (MW) have been known to correlate with dust extinction \citep[e.g.,][]{Merrill34, Hobbs74}. With a high-resolution spectral sample of SNe\,Ia, \cite{Phillips13} argue that the equivalent width of the DIB at $\sim$5780 \AA\ is the most accurate predictor of individual SN extinction. While \NaID absorption can be due to either Na-rich CSM or the interstellar medium (ISM), the DIBs are purely a characteristic of the ISM, suggesting that the dust causing the extinction of SN\,Ia light is predominantly located in the intervening ISM and not in the CSM associated with the SN progenitor.

\begin{figure}
    \centering
    \includegraphics[width=\linewidth]{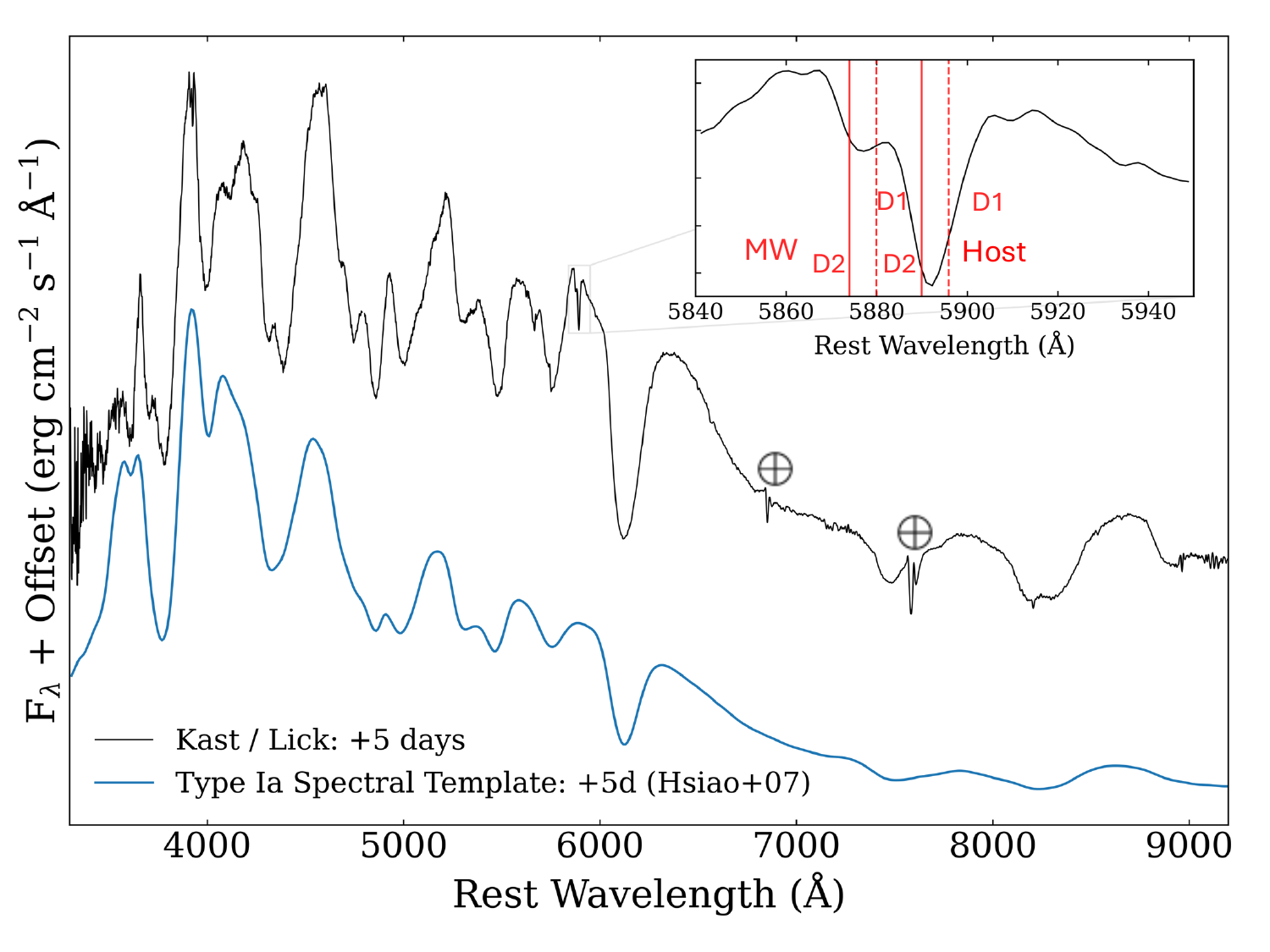}
    \caption{A low-resolution optical spectrum (not corrected for extinction) at 5 days after peak brightness (approximately the midpoint of our high-resolution spectral time-series window) compared with an SN Ia template spectrum at a similar epoch \citep{Hsiao07}. The spectrum is generally consistent with the features expected for a spectroscopically normal SN\,Ia. The inset panel zooms in around Na\,I\,D, where both the MW and host-galaxy components are identified. Telluric features are marked.}
    \label{fig:NaID_lowres}
\end{figure}

SN\,2025rbs (J2000; $\alpha = 22^{\rm hr}37^{\rm m}03.64^{\rm s}$; $\delta = +34^\circ25'07.98''$) was discovered in the nearby galaxy NGC\,7331 and reported to the Transient Name Server\footnote{\url{https://www.wis-tns.org/}} (TNS) by The Gravitational-wave Optical Transient Observer \citep[GOTO;][]{Dyer22} on 2025 July 14 03:22:36 UTC \citep[MJD 60870.14;][]{ONiell25}. We rapidly classified it as a spectroscopically normal SN\,Ia \citep{Andrews25}. In Figure \ref{fig:NaID_lowres}, we present a low-resolution optical spectrum obtained with the Kast spectrograph on the 3\,m Shane telescope at Lick Observatory at +5 days (roughly the midpoint of our high-resolution spectral time-series coverage) that confirms this classification. Unambiguous absorption components associated with both the MW and host-galaxy \NaID are identified. We adopt the redshift of SN\,2025rbs to be identical to its host, NGC 7331 ($z = 0.002722$). This corresponds to a recession velocity of 816 km s$^{-1}$ in the heliocentric reference frame. The distance to NGC\,7331 is constrained through Cepheid anchor measurements at 14.7 $\pm$ 0.6 Mpc \citep{Freedman01}.

In this paper we present and analyze a high-resolution spectral time-series of \rbs at optical wavelengths and a near-infrared (NIR) spectrum at maximum light. We refer to spectral phases with respect to the peak observed brightness of SN\,2025rbs in the $B$ band on 2025 July 28 (MJD 60884) based on an extensive optical photometry campaign to be reported in an upcoming work. Throughout this work, we also report all dates/times in the Coordinated Universal Time (UTC) standard. We present a detailed description of the observed absorption features around Na, Ca, K, and the diagnostic DIB at 5780 \AA. 
\begin{figure*}
    \centering
    \includegraphics[width=\linewidth]{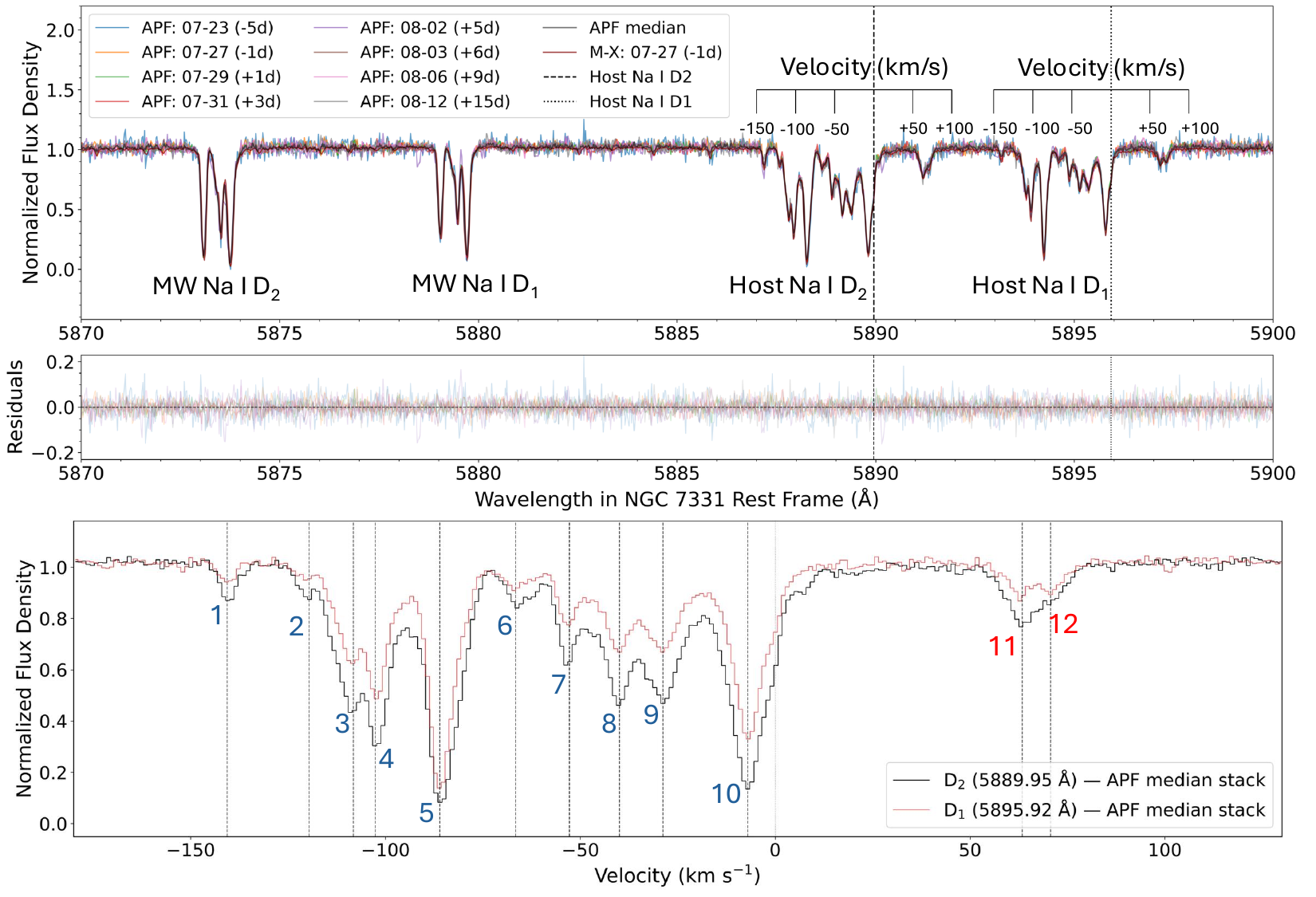}
    \caption{\textit{Top}: High-resolution spectral time-series of SN\,2025rbs with Lick/APF and Gemini/M-X around Na\,I\,D wavelengths between July 23, 2025 and August 12, 2025 ($-$5 to +15 days from peak brightness in the $B$ band). A stacked median of the APF series is also plotted for comparison. Both the D$_{2}$ and D$_{1}$ components associated with MW and the host, NGC\,7331 are marked. The residuals panel shows the deviations of the APF spectra on each night from the median spectrum of the series. Residuals are consistent with noise, confirming lack of time evolution in any of the absorption features. \textit{Bottom}: Zoom-in around Na\,I\,D (D$_{2}$ \& D$_{1}$) in velocity space for the APF median stacked spectrum. For both D$_{2}$ and D$_{1}$, we identify a total of 12 absorption features (10 blueshifted and 2 redshifted) between $-150$ and 100 km s$^{-1}$, assuming their rest-frame wavelengths as zero velocities.}
    \label{fig:NaID}
\end{figure*}

\section{High-resolution Observations of SN\,2025rbs} \label{sec:2}

We obtained 8 high-resolution spectra of SN\,2025rbs between 2025 July 23 and 2025 August 12 ($\sim$20 day window) using the 2.4\,m Automated Planet Finder (APF) at Lick Observatory. The APF hosts the Levy Spectrograph, a high-resolution optical echelle spectrograph \citep[resolution $R \approx 110,000$ at 5500 \AA;][]{Vogt14}. Each APF spectrum covers a wavelength range of 3790--9700 \AA. At $-$1 days from peak brightness, we also obtained a single high-resolution MAROON-X optical spectrum \citep[M-X;][]{Seifahrt18}, as well as a NIR spectrum with IGRINS-2 \citep[][]{Suh25}, both hosted at Gemini-North. M-X and IGRINS-2 offer resolutions of $\sim$80,000 and $\sim$45,000 across wavelength ranges of 5000--9200 \AA\ and 14,900--24,600 \AA, respectively. 

For the APF echelle data, we implement order-by-order wavelength calibrations and a barycentric correction based on Lick coordinates following the detailed description by \citet[][see Section 3.2]{Fulton15}. A detailed DRAGONS-based data-reduction pipeline for M-X echelle spectra, MAROONXDR\footnote{\url{https://github.com/GeminiDRSoftware/MAROONXDR}}, was utilized for the corresponding per-night etalon wavelength solution and computing the barycentric correction. All reduced spectra were then shifted to the rest wavelength of NGC 7331, adopting a redshift of 0.002722. Since APF spectra are calibrated in air while its M-X counterpart in vacuum, we shift the M-X to corresponding air wavelengths for a consistent comparison across our spectral series.

The IGRINS-2 data (NIR H\&K bands) were reduced using version 3.2 of the IGRINS-2 data-reduction pipeline \citep{Sim14,Kaplan24,Sawczynec25}. The software performs the typical reduction steps necessary to process NIR spectroscopic data: combining the frames for each slit position, subtracting the combined B frames from the combined A frames to remove sky emission lines, applying a flat field to the result, rectifying the individual orders, determining a wavelength solution, and extracting the spectrum for each order. The reduced spectra were corrected for telluric absorption and flux-calibrated using the observations of the A0\,V star.  The ends of the individual orders (within H\&K) were then trimmed and merged using the tools available in the Spextool package \citep{Cushing04}. We also removed any data with signal-to-noise ratio (S/N) $< 10$ and/or where the nominal atmospheric transmission for Maunakea (smoothed to the IGRINS-2 resolution) was below 0.75.

Each APF and M-X spectrum is also cross-correlated with H$_{2}$O (for Na) and O$_{2}$ (for K) telluric model lines to scale and divide out absorption produced by Earth's atmosphere at each epoch. We present a few sample telluric corrections in Appendix \ref{sec:6.1}. 

\section{Analysis} \label{sec:3}
\subsection{Sodium: Na\,I\,D} \label{sec:3.1}

In the top panel of Figure \ref{fig:NaID}, we show the time series of the APF and M-X spectra around wavelengths of \NaID\!. We clearly identify \NaIDII and \NaIDI associated with the Milky Way (labeled ``MW") and the host galaxy NGC 7331. The Na absorption lines are not saturated, suggesting moderate extinction due to MW and the host galaxy along the line of sight (see Section \ref{sec:3.2}). For both the host \NaID complexes, we identify 12 components, spanning a total velocity range of $\sim 213$ km s$^{-1}$ (bottom panel of Figure \ref{fig:NaID}; Table \ref{tab:NaID}). We assign zero velocity to the wavelengths of \NaIDII (5889.95 \AA) and \NaIDI (5895.92 \AA) calibrated in air \citep{Morton03} within the rest frame of the host. Based on this formalism, two components are redshifted and ten are blueshifted, with respect to the rest frame of the host (bottom panel of Figure \ref{fig:NaID}). 

SN\,2025rbs is located $\sim 11''$ to the northwest of NGC 7331 \citep[inclination of $\sim 76^\circ$; e.g.,][]{deBlok08, Patra18}. H\,I velocity maps of NGC 7331 show that the north side is approaching while the south side is receding along our line of sight \citep{deBlok08}. Thus, any absorption components of \NaID that are blueshifted from the rest wavelength (and not time-varying) are plausibly due to intervening ISM in the host along our line of sight. Depending on relative motion of the clouds with respect to the SN location, their absorption can be blueshifted (approaching clouds), redshifted (receding clouds), or single/symmetric line profiles with a zero-velocity cloud, all with respect to the velocity of the host galaxy. In SN\,2025rbs, we identify two redshifted components at $\sim 60$--70 km s$^{-1}$ among an otherwise blueshifted set of clouds (Figure \ref{fig:NaID}; Table \ref{tab:NaID}). In addition to a rotating outer disk, NGC 7331 is claimed to have a counterrotating bulge \citep[e.g.,][]{Prada96} with a retrograde rotation component of $\sim$100 km s$^{-1}$ and comprised of molecular gas \citep[e.g.,][]{Tosaki97}. Thus, the mix of components on either side of zero velocity could be a direct probe of the combination of a rotating disk and a counterrotating bulge probing several layers of approaching and receding clouds. 

\begin{figure*}
    \centering
    \includegraphics[width=\linewidth]{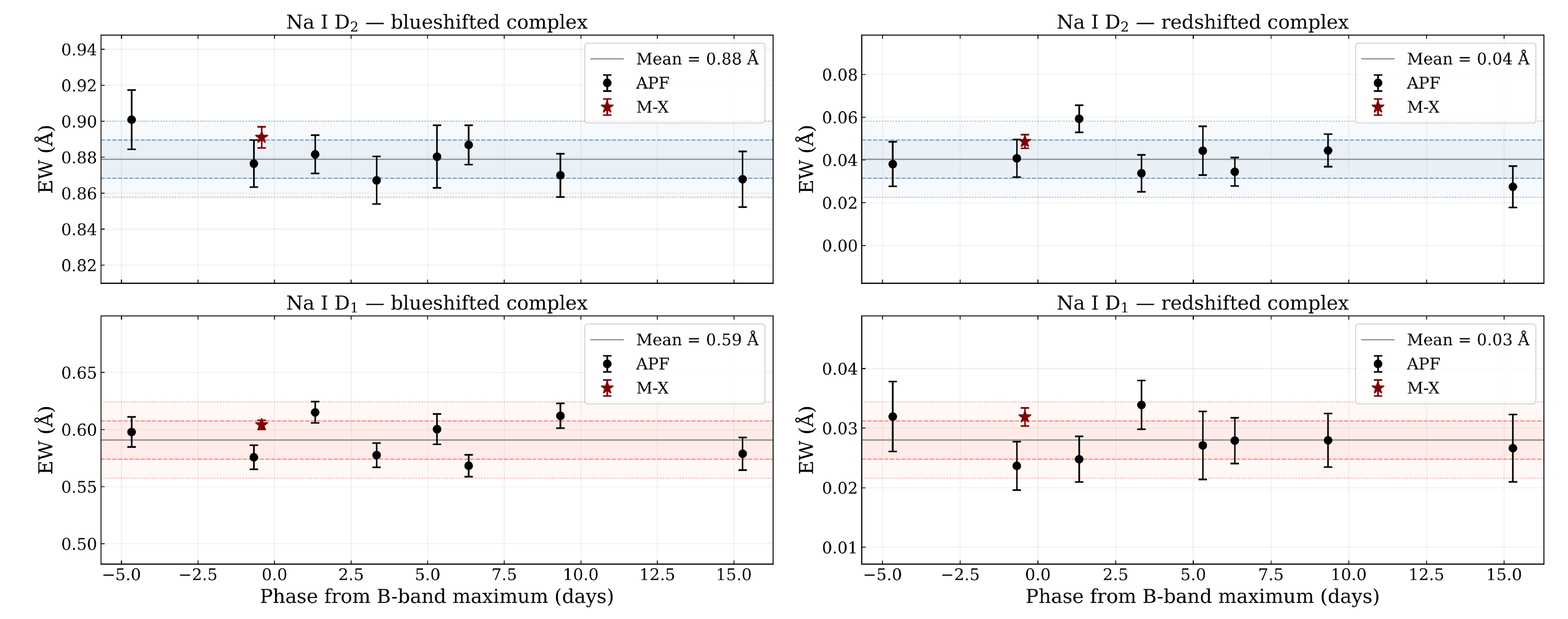}
    \caption{Equivalent-width measurements of \NaIDII and \NaIDI for the blueshifted and redshifted complexes over time. The phase is defined from the epoch of $B$-band maximum brightness on July 28, 2025. The $1\sigma$ and $2\sigma$ scatter is shown as dashed and dotted lines, respectively. All measurements are consistent with each other within $2\sigma$ uncertainties.}
    \label{fig:EW_evolution}
\end{figure*}

\begin{deluxetable}{lcc} 
\tablecaption{\NaID Absorption Lines}
\tablehead{\colhead{Line ID$^{*}$} & \colhead{Wavelengths (\AA)$^{\dag}$} & \colhead{Velocity$^{\ddag}$ (km s$^{-1}$)}}
\startdata
    1  & 5887.19, 5893.16 & $-140.6$ \\
    2  & 5887.59, 5893.56 & $-120.0$ \\
    3  & 5887.83, 5893.79 & $-108.2$ \\
    4  & 5887.93, 5893.90 & $-102.6$ \\
    5  & 5888.26, 5894.22 & $-86.1$  \\
    6  & 5888.65, 5894.62 & $-66.2$  \\
    7  & 5888.91, 5894.88 & $-52.8$  \\
    8  & 5889.16, 5895.13 & $-40.0$  \\
    9  & 5889.38, 5895.35 & $-28.8$  \\
    10 & 5889.81, 5895.78 & $-7.0$   \\
    11 & 5891.19, 5897.17 & $+63.4$  \\
    12 & 5891.32, 5897.29 & $+69.6$
\enddata
\tablenotetext{*}{As marked in Figure \ref{fig:NaID}.}
\tablenotetext{\dag}{\NaIDII and \NaIDI absorption-component wavelengths.}
\tablenotetext{\ddag}{With respect to the rest frame of NGC~7331 (heliocentric radial velocity of 816 km s$^{-1}$).}
\label{tab:NaID}
\end{deluxetable}

An alternate convention in the literature based on high-resolution SN\,Ia samples  is to adopt a zero velocity at the wavelength corresponding to the most intense absorption component \citep[e.g.,][]{Sternberg11, Phillips13}. The zero velocity could also be adopted with respect to narrow emission lines from the host, which probe the rest-frame velocity along the line of sight to the SN position \citep{Maguire13}. When these lines were not identified, \cite{Maguire13} set the rest wavelength using the recession velocity of the host galaxy. This is consistent with our definition of zero velocity. While both approaches have their merits, they do not result in significant differences in the observed statistical preponderance of blueshifted components across the SN\,Ia sample \citep{Phillips13}. 

We find no temporal evolution in any of the absorption components associated with the MW and the host between $-$5 and 15 days with respect to maximum light. The residuals between every APF spectrum and a grand-stack median of all the APF spectra is consistent with noise (Figure \ref{fig:NaID}). We also estimated the integrated equivalent widths (EWs) of the blueshifted and redshifted complexes over time to quantify their evolution (Figure \ref{fig:EW_evolution}). The EWs across all epochs for both complexes are within the 2$\sigma$ uncertainties of each other. Uncertainties in each individual epoch are estimated from the root-mean-square (RMS) of the corresponding spectral continuum. Residuals and integrated EW measurements point to unchanging absorption components, suggesting that all \NaID host-galaxy absorptions are likely caused by clouds along the line of sight in the ISM of NGC 7331.

Blueshifted \NaID absorption with velocities of $-$50 to $-$200 km s$^{-1}$ has been associated with CSM around SNe~Ia, originating as material released from the progenitor system and/or swept up ISM \citep[e.g.,][]{Sternberg11, Maguire13, Phillips13}. Although commonly attributed to the SD system, \cite{Shen13} have shown that blueshifted \NaID  is also a natural byproduct of a He WD companion in the DD scenario. But if these blueshifted absorption lines are due to CSM associated with and surrounding the SN progenitor, over time we would expect their strength to change. In SN\,2025rbs, between $-$5 to +15 days after peak brightness, no single component dominates in absorption strength, and none of them shows any sign of variation. Thus, the most intense component is neither necessarily closest to the site of the SN (in rest-frame velocity) nor associated with its CSM, and is more likely a random ISM cloud along the line of sight with higher density on average than other intervening clouds. Moreover, even if we adopt the alternate convention (zero velocity set at the strongest absorption), SN\,2025rbs still has an asymmetric split of unchanging absorption on either side of the zero velocity. We further discuss properties of these ISM clouds, including their Na column densities, in Section \ref{sec:4.1}.

\begin{figure}
    \centering
    \includegraphics[width=\linewidth]{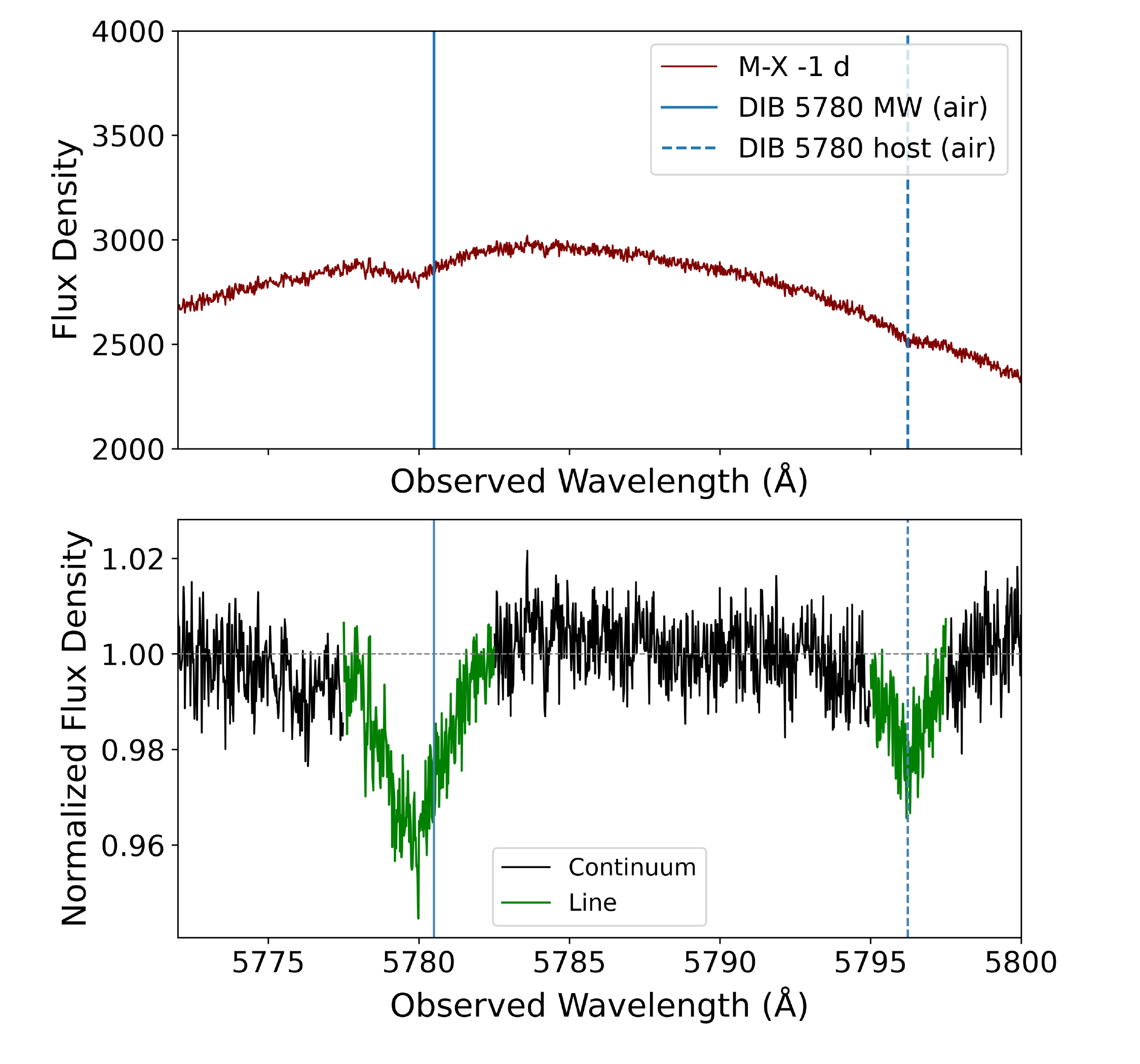}
    \caption{DIB 5780 associated with MW and the host identified in the high S/N M-X spectrum in our spectral time-series. The top panel shows the observed spectrum while the bottom panel is normalized by a fitted continuum model.}
    \label{fig:DIBS_5780}
\end{figure}


\subsection{DIB 5780 and Interstellar Reddening} \label{sec:3.2}
The first DIBs ever detected were in the stellar spectra obtained at Lick Observatory by \cite{Heger22}, and later classified as interstellar spectral absorption lines by \cite{Merrill34}. The sources of these lines still remain only partially understood. An established empirical relation between EW of the \NaID absorption (from low-resolution spectra) and dust extinction as described by \cite{Poznanski12} has typically been used to estimate the extinction of SNe\,Ia. However, \cite{Phillips13} found that $\sim$25\% of SNe\,Ia have stronger \NaID than expected from their extinction ($A_V$), and instead the EW of the DIB at $\sim5780$\,\AA\ has a stronger correlation with the true SN\,Ia dust extinction as, 

\begin{equation} \label{eqn:Av}
\log(EW_{5780}) = 2.283(0.001) + \log A_V \, .    
\end{equation}

This relation has ``a 50\% error in $A_V$ if the 5780\,\AA\ feature is used to estimate the dust extinction for any single object'' \citep[see Equation 6 of][]{Phillips13}, and it is particularly useful as an independent estimate of $A_V$ in cases where the \NaID line is saturated. We identified two absorption components likely associated with DIB 5780 (for MW and the host) in the high S/N M-X spectrum at day -1 (Figure \ref{fig:DIBS_5780}). From direct integration of the absorption profiles in the continuum-normalized spectrum, we estimate a total $EW_{5780} = 122.9\,\pm\,2.7$ m\AA. Plugging this into Equation \ref{eqn:Av}, we derive $A_V = 0.64\,\pm\,0.01 $\,mag (statistical uncertainty). Considering a systematic uncertainty of 50\% dominates over the statistical uncertainties from the EW measurement, the total extinction is $A_V = 0.64\,\pm\,0.32 $\,mag, translating to a reddening of $E(B-V) = 0.21 \pm 0.10 $\,mag (assuming $R_V = 3.1$). Since nightly APF spectra have significantly poorer S/N, we median combine the spectra through stacking and find a consistent $EW_{5780}$ (within uncertainties) as a sanity check. Since our choices are between a single M-X spectrum and the stacked APF spectrum, we cannot search for any potential evolution in the DIB at 5780\,\AA\ from our data.

\begin{figure}
    \centering
    \includegraphics[width=\linewidth]{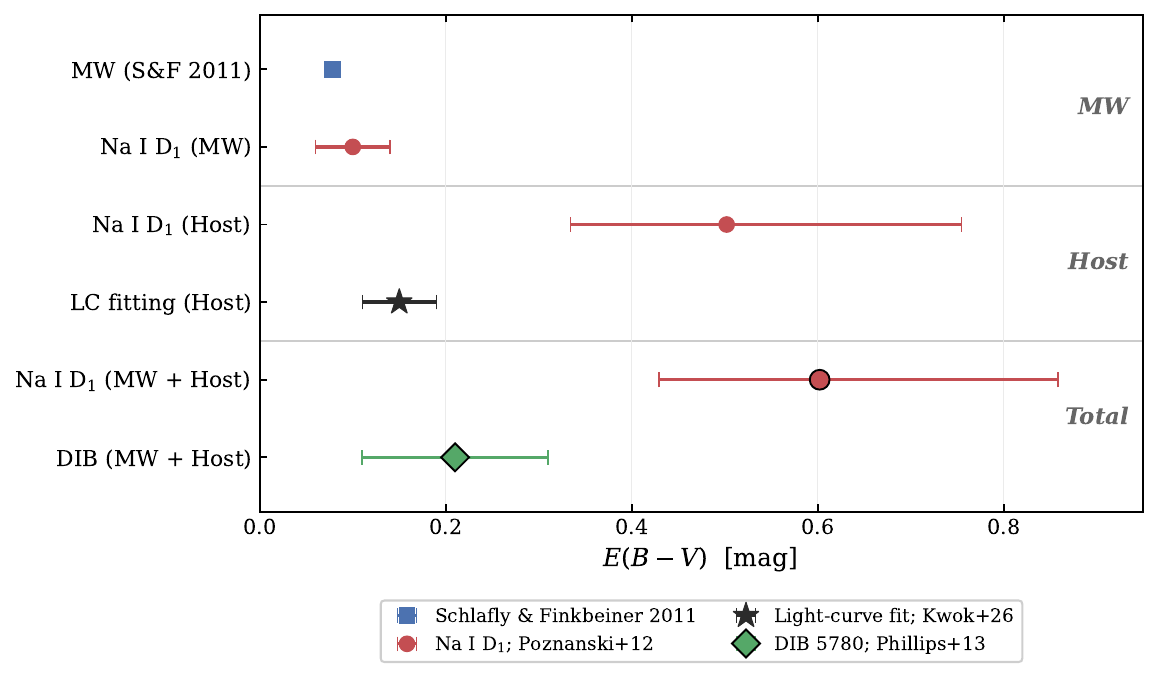}
    \caption{Comparison between interstellar reddening toward SN\,2025rbs inferred from independent methods. Total reddening estimated from DIB 5780 (green rhombus) is consistent with an independent inference from light curve fitting \citep[black star;][]{Kwok26}. MW reddening in the direction of SN\,2025rbs is consistent between dust maps of \cite{Schlafly_Finkbeiner11} (blue square) and from EW of \NaIDI MW absorptions \citep[red circle; based on empirical relations of][]{Poznanski12}. We find that the observed \NaIDI strength in SN\,2025rbs due to host absorption significantly overestimates the line of sight reddening (compared to more accurate tracers like DIB 5780) as previously established by \cite{Phillips13} for $\sim$25\,\% of SNe\,Ia.}
    \label{fig:reddening_comparison}
\end{figure}


In Figure \ref{fig:reddening_comparison}, we compare the reddening estimates from several independent methods against our DIB-based $E(B-V)$. The total reddening (MW + Host) estimated with DIB 5780 is consistent with a moderate amount of interstellar reddening, $E(B-V)_\mathrm{LC} = 0.15\,\pm\,0.04 $\,mag, inferred from independent light-curve fitting of SN\,2025rbs \citep[Figure \ref{fig:reddening_comparison};][]{Kwok26}. Reddening due to MW in the direction of SN\,2025rbs based on the Galactic dust maps of \cite{Schlafly_Finkbeiner11} is $E(B-V)_{MW} = 0.078\,\pm\,0.001$ mag. The direct integration of \NaID absorptions due to MW (Figure \ref{fig:NaID}) gives $EW_\mathrm{{D_{1}}}$ = 332.2 $\pm$ 11.1 m\AA, which can be translated to a reddening of $E(B-V)_\mathrm{MW} = 0.09\,\pm\,0.04$ mag using the empirical relations of \cite{Poznanski12}. Thus, line of sight extinction due to MW is commensurate with the observed strength of MW \NaID lines. On the other hand, for the host \NaID components, the $EW_\mathrm{{D_{1}}}$ = 0.618 $\pm$ 0.020 \AA\, translates to a much larger reddening of $E(B-V)_\mathrm{Host} = 0.50 ^{+0.25}_{-0.17} $ mag \citep[][]{Poznanski12} (Figure \ref{fig:reddening_comparison}), suggesting an anomalously strong \NaIDI absorption feature in comparison to the true extinction. Considering SN\,2025rbs has a preponderance of blueshifted features (Section \ref{sec:3.1}), presence of anomalously strong Na absorptions is remarkably consistent with the findings of \cite{Phillips13}. This further strengthens our interpretation that the dust causing the extinction of light from SN\,2025rbs is likely predominantly located in the intervening ISM and not in the CSM associated with the SN progenitor.


\subsection{Calcium: Ca\,II (H\&K)} \label{sec:3.3}
\begin{figure}
    \centering
    \includegraphics[width=0.5\textwidth]{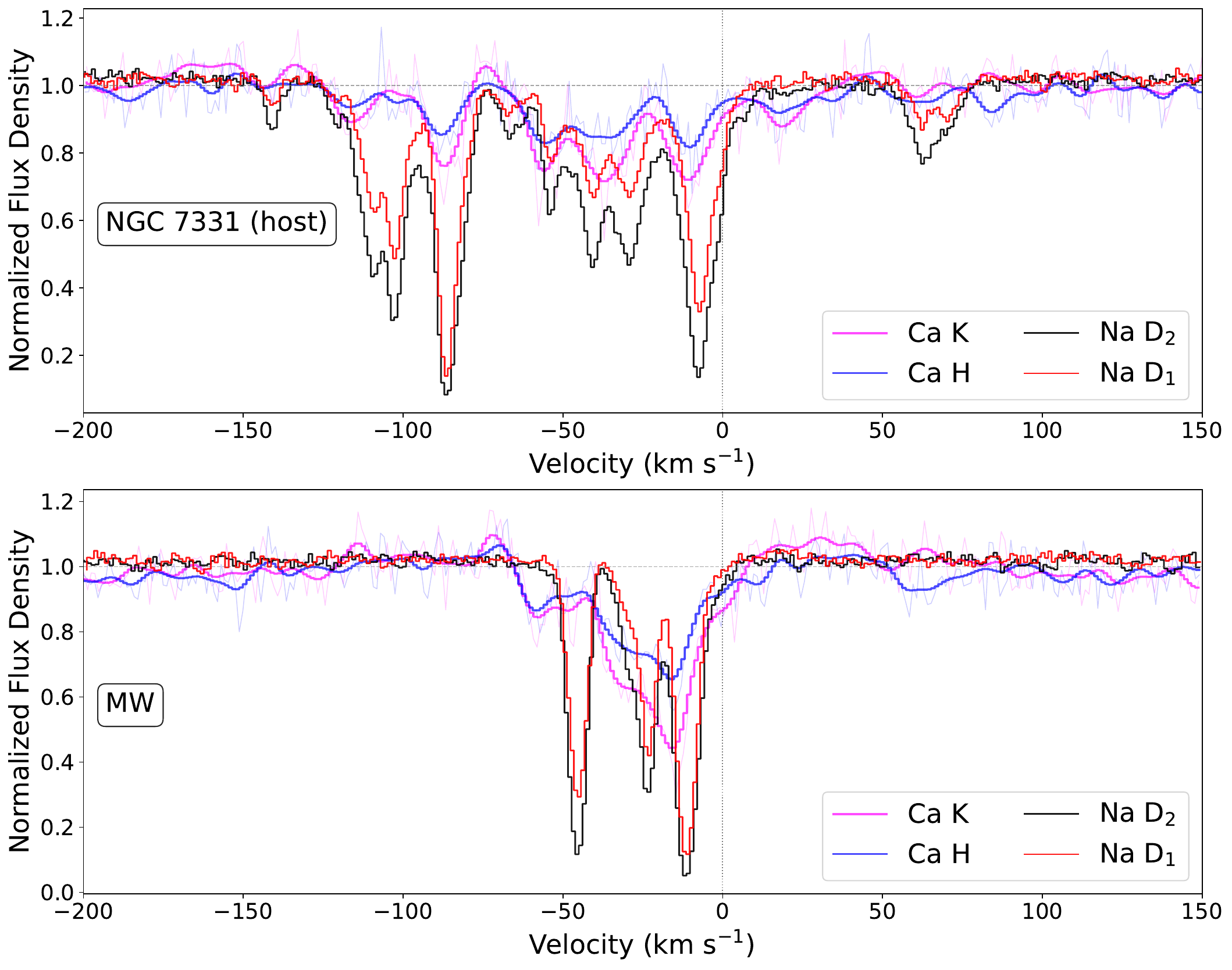}
    \caption{Comparison of continuum-normalized Ca\,II (H\&K) profiles in velocity space with corresponding \NaIDII and \NaIDI profiles for both the host NGC 7331 (top) and MW (bottom). Zero velocity corresponds to the rest wavelength of each line. The Ca\,II spectra have been smoothed (bin size of 3\,km s$^{-1}$) for presentation. The general agreement between the velocity distributions of some of the individual absorption components of Ca and Na suggest a common origin for both elements in the same intervening clouds.}
    \label{fig:CaII}
\end{figure}

We identify absorption lines due to Ca\,II (H\&K) in the APF spectral time series. The stacked median spectrum is analyzed, given the low sensitivity of APF and the lack of M-X coverage in this wavelength range. Previous literature studies that identified time-varying \NaID absorption features did not note any evolution in the Ca\,II (H\&K) owing to a larger ionization potential \citep[e.g.,][]{Patat07, Simon09}. 

Adopting a similar procedure as that for \NaID and the DIB 5780, we estimate an integrated EW of $\sim$0.6\,\AA\ and $\sim$0.4\,\AA\ for Ca\,II H and Ca\,II K absorptions within NGC 7331, respectively, from their continuum-normalized spectra. In Figure \ref{fig:CaII}, we plot the velocity profile of Ca\,II H and Ca\,II K in comparison to \NaIDII and \NaIDI for both the host galaxy and MW. Several Ca absorption lines trace the Na gas, suggesting a common origin for the absorbing Ca and Na gas at those velocities, although the S/N and sensitivity around Ca\,II H and Ca\,II K is signficantly lower. 

\begin{figure}
    \centering
    \includegraphics[width=0.5\textwidth]{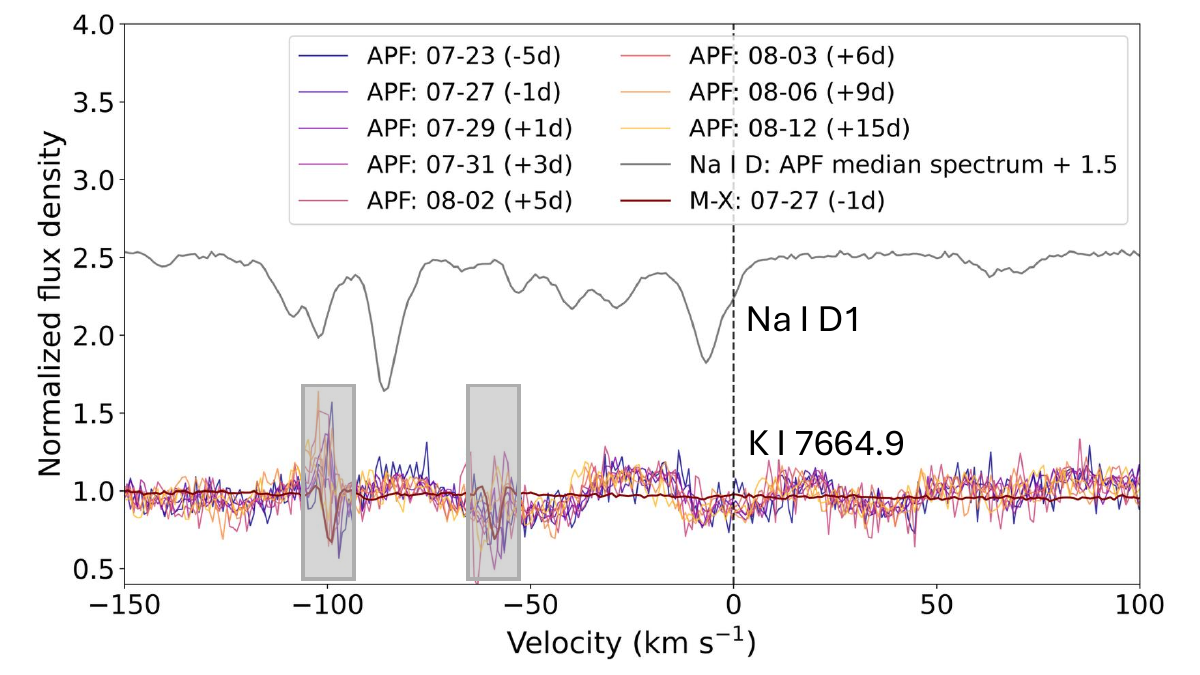}
    \caption{Comparison of K\,I 7664.9 \AA\ in velocity space with \NaIDI APF median profile. The second component of K\,I falls between the APF orders. The gray boxes indicate residual contributions from telluric absorption. Compared to the flat M-X spectrum, the APF spectra suffer from significant fringing. No clear absorption associated with K\,I is apparent.}
    \label{fig:KI}
\end{figure}

\subsection{Potassium: K\,I} \label{sec:3.4}

\begin{figure*}
    \centering
    \includegraphics[scale=0.4]{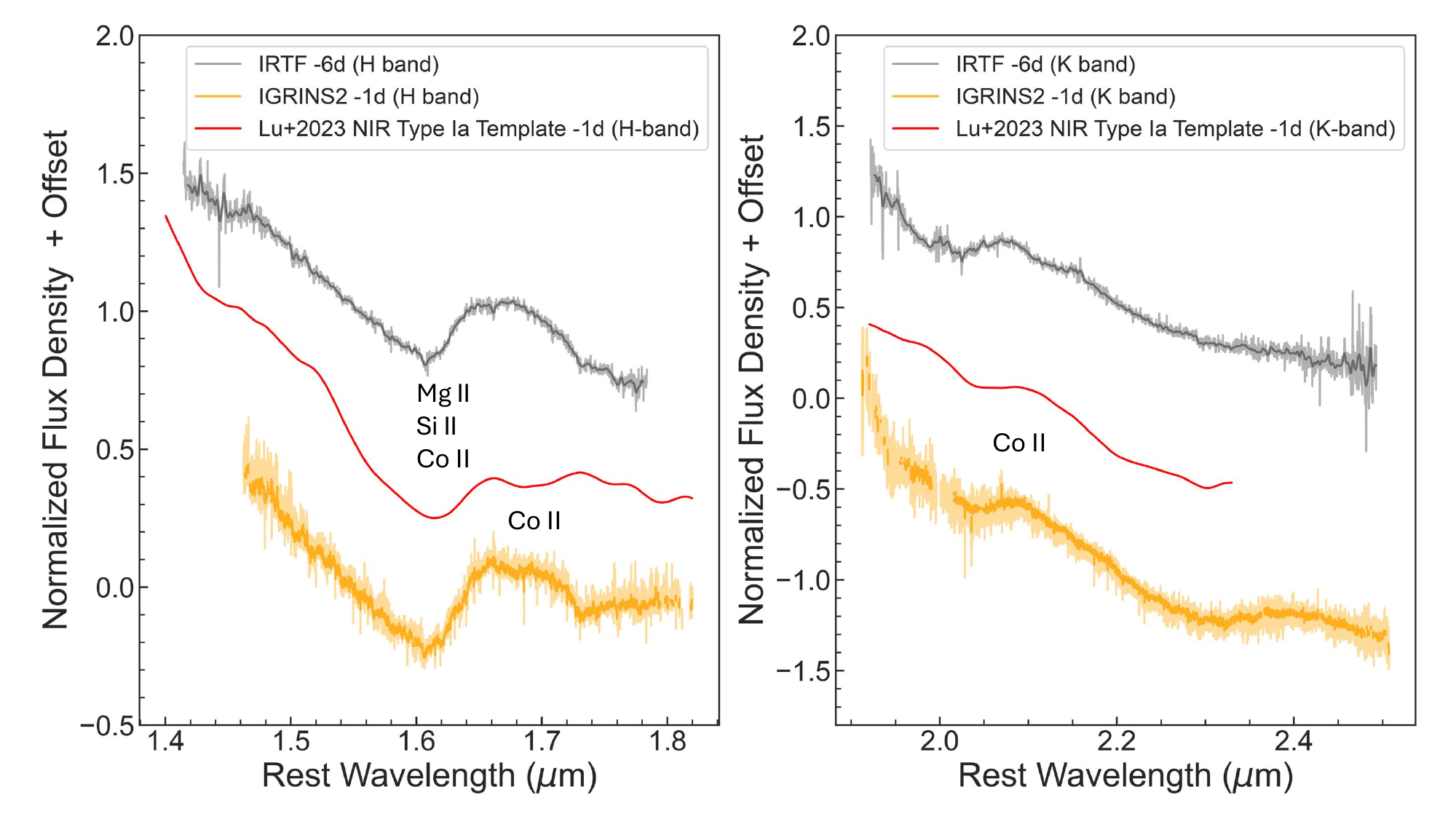}
    \caption{Comparison of the order-stitched and telluric-corrected IGRINS-2 spectrum at $-$1 days from $B$-band maximum with a low-resolution IRTF spectrum in the H\&K bands. Both spectra agree in broad shapes and features. The rising spectral feature intrinsic to the SN at $\sim$1.6\,$\mu$m observed independently in both spectra is reminiscent of the $H$-band break typically seen in SN\,Ia spectra. A NIR template spectrum of a typical SN\,Ia at $-$1 days from \cite{Lu23} is shown for comparison.}
    \label{fig:NIR}
\end{figure*}

The evolution of the K\,I doublet (7664.90 and 7698.96\,\AA) is an independent probe of contributions from CSM around the SN progenitor \citep[e.g.,][]{Graham15}. We present the time series of \KI 7698.96\,\AA\ with the \NaIDI profile overlaid in the same velocity space in Figure \ref{fig:KI}. Zero velocity of \NaIDI and \KI correspond to their wavelengths in the rest frame of the host galaxy. No temporal evolution is noted in any component. Unfortunately, the second \KI component of the doublet falls between the APF orders. Additionally, significant fringing is observed in the APF spectra, making the clear identification of any component produced by \KI difficult. We also plot the M-X spectrum at peak brightness (with significantly weaker fringing), showing a flat spectrum in the wavelength range. This confirms that the bumpy continua observed in all the APF spectra are not real spectral features. We discuss the expected column densities of \KI from observed components of \NaID based on the empirical relations of \cite{Phillips13} and compare with our observations in Section \ref{sec:4.1}.

\begin{figure*}
    \centering
    \includegraphics[width=\textwidth]{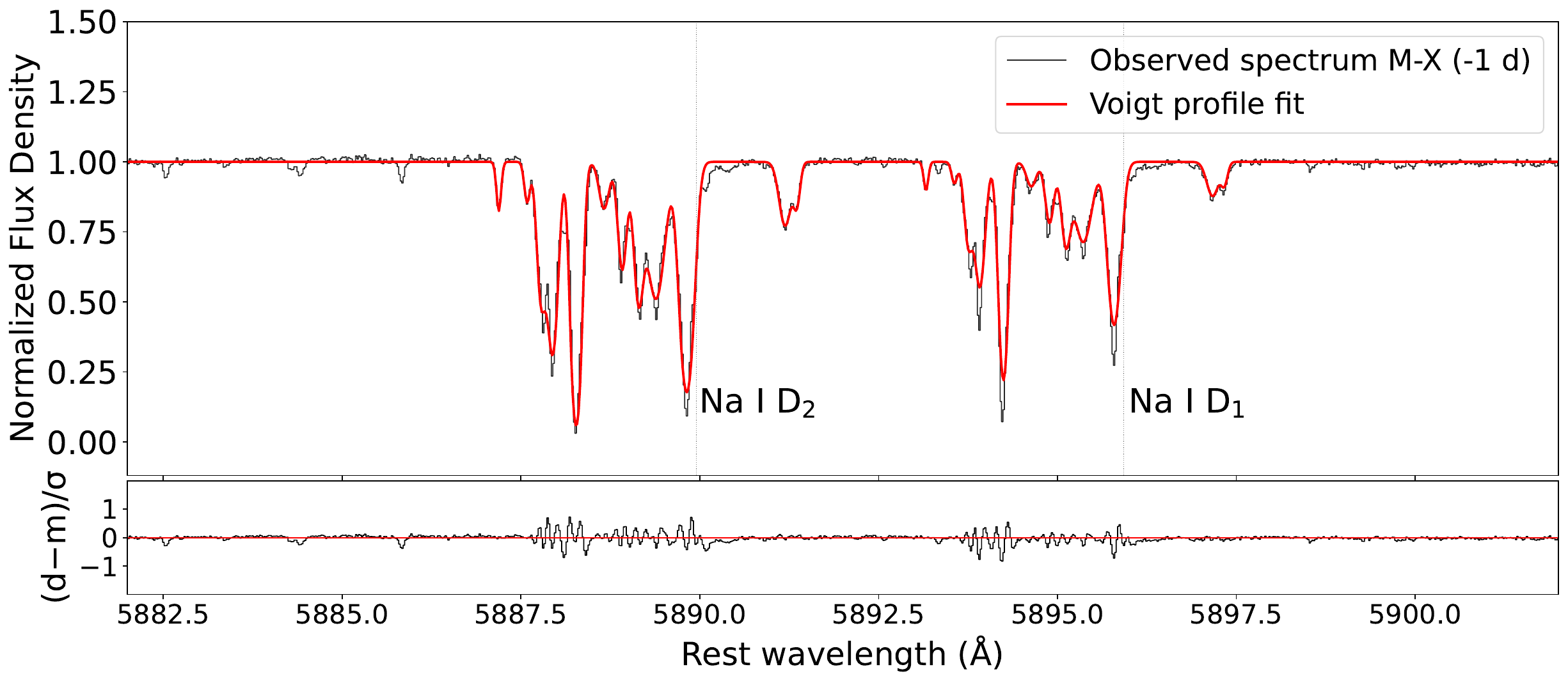}
    \caption{Simultaneous Voigt profile fitting of the host-galaxy \NaIDII and \NaIDI in the high-S/N continuum-normalized M-X spectrum. The dotted lines represent the rest wavelengths of \NaIDII and \NaIDI in the host reference frame. Most absorption components are correctly identified and the column density associated with each cloud is estimated. In the bottom panel, residuals between the data and model normalized by noise in the data are plotted.}
    \label{fig:NaID_Voigt}
\end{figure*}

\subsection{Near-infrared Spectrum around Peak Brightness} \label{sec:3.4}

High-resolution spectra of Type Ia SNe at NIR wavelengths are quite rare, but can serve as an independent pathway to detect and track narrow absorption features in searching for CSM signatures and for quantifying line-of-sight extinction. In this work we present a single IGRINS-2 spectrum of SN\,2025rbs near peak brightness. Outside of telluric absorption, no narrow features can be reliably identified in any of the IGRINS-2 orders. Thus, we show the stitched (across all echelle orders) and telluric-corrected IGRINS-2 high-resolution spectrum (see Section \ref{sec:2}) of SN\,2025rbs at $-$1 days from $B$-band peak brightness in the H and K bands (Figure \ref{fig:NIR}). For comparison, we plot a low-resolution NIR spectrum obtained with IRTF at $-6$ days and a NIR template of an SN\,Ia at $-$1 days \citep{Lu23}. All three agree in broad shape and features; this consistency between an independent low-resolution spectrum and the order-stitched IGRINS-2 data indicates that the observed features are intrinsic to the SN rather than artifacts of order stitching or telluric correction. 

Around 1.6\,$\mu$m, both spectra show a rising H-band flux typically observed in SNe\,Ia as the photosphere recedes into the $^{56}$Ni-rich ejecta and multiplets of Fe\,II/Co\,II/Ni\,II are exposed, blanketing the blue side and re-emitting redward \citep[e.g.,][]{Kirshner73, Wheeler98, Hsiao13}. The typically observed ``$H$-band break'' in SN\,Ia spectra are expected to appear soon past maximum light ($\sim$3 days) and rise to a peak intensity at $\sim 10$--12 days \citep{Hsiao13}, so we attribute the observed feature between $-$6 and $-$1 days in Figure \ref{fig:NIR} to be more likely a photospheric intermediate-mass element line blend at these pre-maximum epochs. These features are all within expectations in the NIR for SNe\,Ia \citep{Marion09}.

\section{Discussion} \label{sec:4}

Our high-resolution spectral series of SN\,2025rbs covers between $-$5 and 15 days from peak brightness in the $B$ band. The last nondetection of SN\,2025rbs is well constrained ($\sim$1 day before discovery). Assuming the explosion epoch to be the midpoint between first detection (MJD 60870.14) and last nondetection (MJD 60869.36) at MJD 60869.75 (July 13, 2025), our high-resolution coverage extends 10--30 days after explosion. Our data conclusively shows a lack of temporal evolution in any of the absorption components, hinting at their ISM origin. Since these absorption components are resolved in our high-resolution data, we can estimate the total column density of gas associated with different species (e.g., Na, Ca) to make inferences about the properties of the absorbing ISM gas and discuss CSM constraints around the progenitor of SN\,2025rbs. 

\subsection{Absorbing Gas Column Densities} \label{sec:4.1}

We estimate the column densities of the host absorption components associated with the \NaID doublet using Voigt profile fitting with the minimum velocity, $v_{o, i}$ (in km s$^{-1}$), column density, $\log (N_{i}/$cm$^{-2})$, and the Doppler parameter, $b_{i}$ (in km s$^{-1}$) associated with each absorption cloud $i$ as free parameters. Absorption components are identified by smoothing the median-stacked APF and the M-X spectra with a Gaussian kernel and ensuring the peak depth is at least three times stronger than the underlying continuum independently across the \NaIDII and \NaIDI wavelength ranges. Column densities are estimated by assuming the optically thin regime. We identify 12 absorption components between $-$150 and +100 km s$^{-1}$ across the velocity space for the \NaID doublet. The absorption component velocities derived here are congruent with the labels in the bottom panel Figure \ref{fig:NaID} and Table \ref{tab:NaID}. Atomic parameters are adapted from \cite{Morton03} and the 2:1 oscillator-strength ratio between \NaIDII and \NaIDI is assumed in the optically thin regime \citep{Welty_Hobbs_Kulkarni94}. Both \NaIDII and \NaIDI components are thus fit simultaneously.

We present our fitting results for the higher S/N M-X spectrum in Figure \ref{fig:NaID_Voigt}. As an independent test, similar fits to the median stacked APF spectra around \NaID absorptions give consistent results within uncertainties. We also fit the observed Ca\,II (H\&K) absorption from the stacked APF median spectrum, albeit at worse S/N compared to \NaID (no M-X coverage at this wavelength). Since each absorption cloud is along the line of sight, the total column density is the addition of all identified individual components. We estimate total column densities of the order $\log (N_\mathrm{Na\,I}\,$cm$^{-2}$$)\approx$ 12.9 and $\log (N_\mathrm{Ca\,II}\,$cm$^{-2}$$) \approx$ 12.8, for Na and Ca species along the line of sight within the host of SN\,2025rbs, respectively. Voigt fit parameters associated with individual absorption components of Na and Ca are presented in Appendix \ref{sec:6.2} (Table \ref{tab:Voigt_NaID} \& Table \ref{tab:Voigt_CaII}). Photon absorption from galactic halo gas is typically expected to have larger column densities of Ca\,II (H\&K) relative to \NaIDnospace, whereas for ISM/CSM clouds, the ratios of column densities are of order unity \citep{Baldwin85}. Thus, with  $\log (N_\mathrm{Na\,I}$\,/\,$N_\mathrm{Ca\,II}) \approx 0.1$, we can strongly exclude the possibility of a galactic halo gas origin for the line-of-sight absorption.

With a large and heterogeneous sample of SN spectra, larger columns of ISM column density were correlated with environments that are more massive, more actively star-forming, younger, and viewed from a more inclined angle \cite{Gonzalez-Gaitan25}. The host of SN\,2025rbs, NGC 7331 is a massive, unbarred SA(s)b spiral frequently described as a structural analog of the Milky Way, viewed at high inclination \citep[inclination of $\sim 76^\circ$; e.g.,][]{deBlok08, Patra18} and hosting a prominent circumnuclear star-forming ring \citep[e.g.,][]{Telesco82, Smith04} contributing to over one-third of its total active star-formation \citep{Thilker07}. This offers additional explanations for the observed strong column and unchanging Na I D absorptions along the line of sight to SN 2025rbs and reinforces our interpretation of the anomalous \NaID as disturbed/diffuse host ISM rather than CSM.

\begin{figure}
    \centering
    \includegraphics[width=0.5\textwidth]{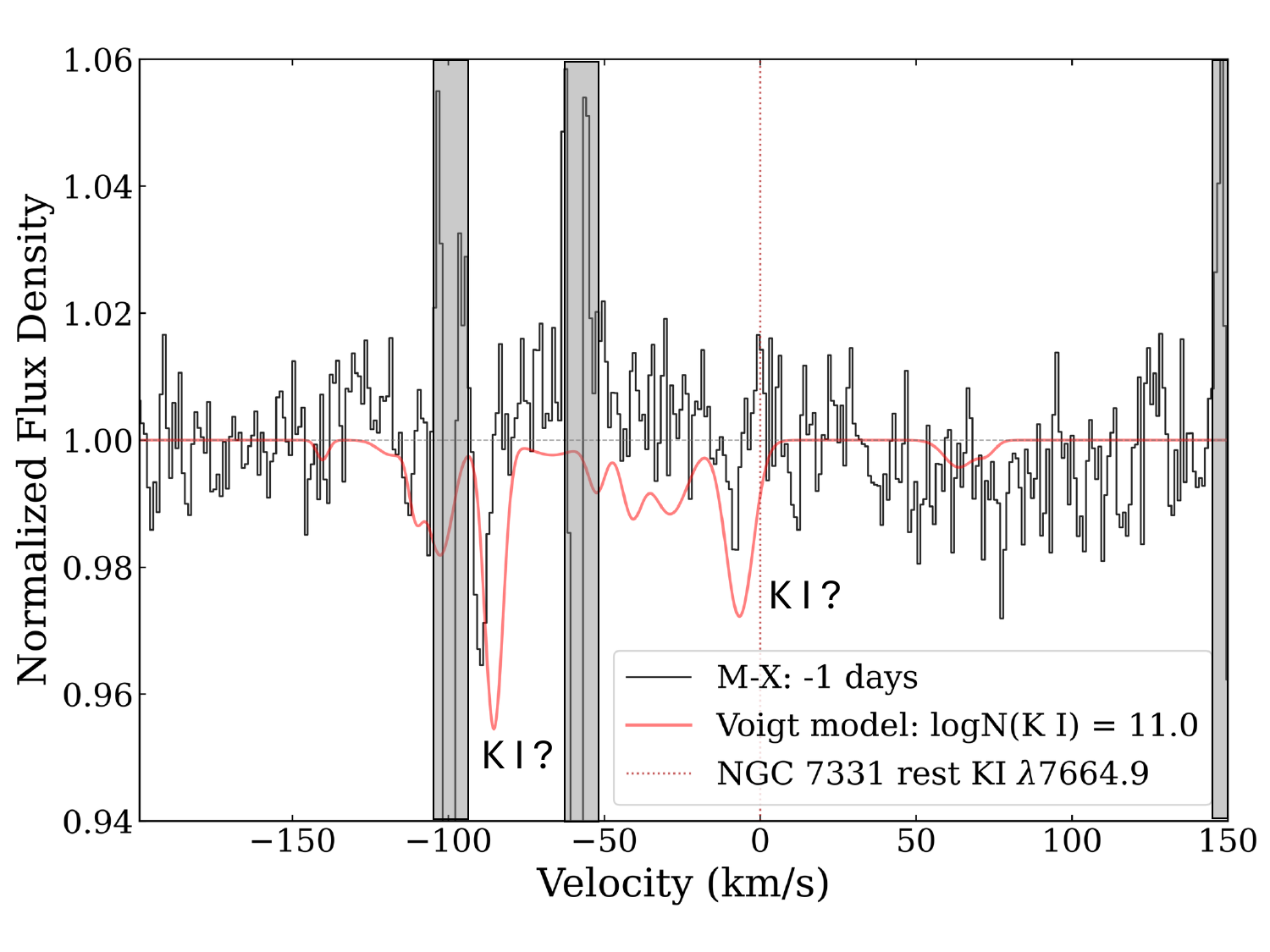}
    \caption{Comparison of continuum-normalized K\,I 7664.9\,\AA\ in the M-X spectrum with a scaled Voigt model assuming a column density of $\log N_\mathrm{K\,I} =$ 11, a value based on the estimated $\log N_\mathrm{Na\,I}$ and the empirical relationship presented by \cite{Phillips13}. Common velocities between K and Na components are assumed in the construction of the K\,I Voigt model. Gray vertical regions mask residuals from telluric absorption. Two tenuous absorption lines in the observed spectrum could be due to K\,I, although the low S/N does not allow us to confirm this.}
    \label{fig:KI_Voigt}
\end{figure}

As shown in Figure \ref{fig:KI}, the S/N of our spectra prohibits any detailed analysis of K\,I absorption in SN\,2025rbs. Based on the empirical column-density relationship between Na\,I and K\,I from the SN\,Ia sample of \cite{Phillips13}, for a column density of $\log (N_\mathrm{Na\,I}\,$cm$^{-2}$$) \approx$ 12.9, the corresponding expected $\log (N_\mathrm{K\,I}\,$cm$^{-2}$$)$ is $\approx$ 11. In an ideal scenario, if we assume the same cloud origin for each species (i.e., similar velocity distributions for each absorption component), we can scale the best-fit \NaID Voigt model to a new column density of 11 dex to represent the expected profile from \KI. With nearly identical $f$-values between the alkali resonance doublets Na\,I and K\,I \citep{Morton03} and at a column density of 11 dex, the expected absorption depth will be $\sim10^{2}$ times weaker, since optical depth scales as column density (in the optically thin regime). We compare the scaled model with the observed K\,I spectrum in Figure \ref{fig:KI_Voigt}. Two potential absorption lines are marked in the spectrum that could be due to K\,I. Identification of any absorption component with respect to model K\,I absorption is tenuous at best and cannot be confirmed at our observed S/N. Moreover, one of the absorption lines is close to a region marred by a telluric band, adding to the uncertainty of its identification. 

\subsection{CSM Constraints} \label{sec:4.2}

Our estimates of Na\,I column density in Section \ref{sec:4.1} can be used to approximate the hydrogen column density along the line of sight to SN\,2025rbs. The column density of neutral sodium can be expressed as $N_{\mathrm{Na}}$ = $N_{\mathrm{Na\,I}}/X = 7.9 \times 10^{12}/X$ cm$^{-2}$, where $X = N_\mathrm{Na\,I}/N_\mathrm{Na}$ is the Na ionization fraction.  Assuming solar Na abundance of 12 + $\log$(Na/H) = 6.17 \citep{Asplund05}, we can estimate Na/H = 1.5 $\times$ 10$^{-6}$. Rewriting $N_{\mathrm{H}} = N_{\mathrm{Na}}/(\mathrm{Na/H})$ and plugging in the values, we estimate $N_{\mathrm{H}} = 5.3\times10^{18}/X$\,cm$^{-2}$.  If we assume a SN\,2006X-like CSM around the progenitor within a thin shell with a radius of 10$^{16}$ cm \citep{Patat07}, the corresponding mass would be 5.6$\times$10$^{-6}/X$\,M$_{\odot}$. However, since no absorption component changes with time (unlike in SN\,2006X), this gas is likely associated with ISM in NGC\,7331 (along the line of sight) and not CSM from the vicinity of the SN progenitor. Thus, lack of any temporal evolution in the observed absorption features disfavors such a geometry of CSM around the progenitor of SN\,2025rbs.



Additionally, closer to the SN site, there is freely expanding ejecta from the explosion itself. For a characteristic outer-ejecta velocity of $\sim$10,000--20,000 km s$^{-1}$, the ejecta reach $r\approx(3$--$5)\times10^{15}$\,cm by the end of our coverage (30 d after explosion).  Dense CSM within this radius would be shocked, producing signatures absent in a clean SN\,Ia: intermediate-width or broad H$\alpha$ and other Balmer/He emission from the shock, narrow recombination lines from the unshocked photoionized gas, a blue quasi-continuum excess, and/or radio/X-ray emission, as in the SN\,Ia-CSM events SN 2002ic and PTF11kx \citep[e.g.,][]{Hamuy03, Dilday12, Silverman13}. While most SNe\,Ia with time-varying \NaID absorptions \citep[barring a few exceptions e.g., PTF11kx][]{Dilday12} have been spectroscopically normal, the converse is not generally true. Several spectroscopically normal SNe\,Ia show no time evolution in multi-epoch high-resolution observations \citep[e.g.,][]{Sternberg14}. SN\,2025rbs is also a photometrically and spectroscopically normal SN\,Ia, with no narrow or intermediate-width emission, no H$\alpha$, and no interaction-driven flux excess during our coverage. We therefore conclude that there is no evidence for dense CSM within $\sim 10^{15}$ cm of the progenitor. 

As all observed absorption components identified in SN\,2025rbs seem to have an ISM origin, strong progenitor system constraints cannot be derived from our data alone. That said, lack of dense CSM close to the progenitor plausibly disfavors the SD-wind signature — leaving the observations more aligned with (though not uniquely proving) a DD origin involving CO WDs \citep[e.g.,][]{Maoz14}. However, it is to be noted that in case of a degenerate He-CO WD companion, some CSM pollution of the WD environment is still expected in the DD scenario \citep{Shen13}. Additionally, our limits of non-detection only applies to CSM dense enough to drive detectable interaction and only out to the radius reached by the ejecta during our observations; low-density, H-poor, or out-of-sightline material would not necessarily be revealed.



\section{Summary \& Conclusions} \label{sec:5}

We have presented a high-resolution optical spectral time-series of SN\,2025rbs, obtained between $-5$ and $+15$\,days relative to $B$-band maximum light. At optical wavelengths, our eight APF epochs and one M-X epoch resolve the intervening \NaIDnospace,
Ca\,II~(H\&K), and tenuous \KI absorption components along the sightline to a spectroscopically normal SN\,Ia in the nearby galaxy NGC\,7331. Our principal results are as
follows:

\begin{enumerate}

\item We resolve 12 host-galaxy \NaID components (10 blueshifted, 2 redshifted)
spanning $\sim$213\,km\,s$^{-1}$ (Figure~\ref{fig:NaID}, Table~\ref{tab:NaID}).
All components are unsaturated and consistent with the moderate line-of-sight
reddening.

\item None of the resolved components shows evidence of temporal variability.
The per-epoch residuals about the median APF spectrum are consistent with noise,
and the integrated EWs of both the blueshifted and redshifted complexes
are constant within 2$\sigma$ over our coverage (Figure~\ref{fig:EW_evolution}).

\item From the 5780\,\AA\ DIB (MW + Host) in the high-S/N M-X
spectrum (EW$_{5780} = 122.9 \pm 2.9$\,m\AA), we infer a line-of-sight extinction
$A_V = 0.64 \pm 0.32 $\,mag, or $E(B-V) = 0.21\,\pm\,0.10$\,mag, assuming $R_V=3.1$ \citep{Phillips13}. This value of reddening is consistent with an independent estimate from the optical light-curve fitting \citep{Kwok26}. Anomalously strong \NaID in comparison to true extinction (from DIB 5780) is observed in the host of SN\,2025rbs, suggesting most of this extinction likely comes from the ISM in NGC 7331.

\item Simultaneous Voigt-profile fitting yields total host-galaxy column densities of
$\log(N_\mathrm{Na\,I}\,$cm$^{-2}$$) \approx 12.9$ and $\log(N_\mathrm{Ca\,II}\,$cm$^{-2}$$) \approx 12.8$ for Na and Ca, respectively (Figure~\ref{fig:NaID_Voigt}, Table~\ref{tab:Voigt_NaID}, and Table~\ref{tab:Voigt_CaII}). The Ca\,II
components generally trace \NaID in velocity (Figure~\ref{fig:CaII}), indicating a common
origin at those velocities. An order-unity column density ratio of $\log(N_\mathrm{Na\,I}/N_\mathrm{Ca\,II}) \approx 0.1$ excludes a galactic-halo origin in favor of absorption in the disk ISM of NGC\,7331 \citep{Baldwin85}.

\item Taken together, the temporal invariance of all resolved \NaID
components, combined with the absence of Balmer emission, narrow or
intermediate-width emission, and any interaction-driven continuum excess, reveals no evidence for CSM around the progenitor of \rbs\!. While the ISM components cannot place strong constraints on the progenitor system, lack of detectable CSM features point to a DD configuration, although line of sight cannot completely rule out a non-degenerate companion. CSM non-detection limits
apply only to gas dense enough to produce detectable ejecta-CSM interaction, lying along the sightline and within the radius reached by the ejecta
($r\approx(3$--$5)\times10^{15}$\,cm) over our $\sim$10--30\,day post-explosion
coverage. Low-density, H-poor, or out-of-sightline material cannot not excluded.

\end{enumerate}

\noindent \rbs\!, thus joins the majority of spectroscopically normal SNe\,Ia
that exhibit no evidence for circumstellar gas, while its fully resolved sightline
demonstrates the power of few km\,s$^{-1}$ time-series spectroscopy for
separating host-galaxy ISM from progenitor CSM in nearby bright SNe.

\begin{acknowledgments}


S.V. and the UC Davis time-domain research team acknowledge support from National Science Foundation (NSF) grant AST-2407565. J.E.A. is supported by the international Gemini Observatory, a
program of NSF's NOIRLab, which is managed by the Association of Universities for Research in Astronomy (AURA) under a cooperative agreement with the NSF, on behalf of the Gemini partnership of Argentina, Brazil, Canada, Chile, the Republic of Korea, and the United States of America. 
Time-domain research by the University of Arizona team and D.J.S. is supported by NSF grants 2308181, 2407566, and 2432036. 
Supernova research at Rutgers University is supported in part by NSF award AST-2407567. S.W.J. is grateful for a Guggenheim Fellowship.
A.V.F.’s research group at UC Berkeley acknowledges financial assistance from the Christopher R. Redlich 
Fund, Gary and Cynthia Bengier, Clark and Sharon Winslow, Alan Eustace and Kathy Kwan (W.Z. is a Bengier-Winslow-Eustace Specialist in Astronomy), Timothy and Melissa Draper, Briggs and Kathleen Wood, Ellyn and Alan Seelenfreund (T.G.B. is Draper-Wood-Seelenfreund Specialist in Astronomy), and numerous other donors. N.F. acknowledges support from the NSF Graduate Research Fellowship Program under grant DGE-2137419.

A major upgrade of the Kast spectrograph on the Shane 3\,m telescope at Lick Observatory, led by Brad Holden, was made possible through gifts from the Heising-Simons Foundation,  
William and Marina Kast, and the University of California Observatories. Research at Lick Observatory is partially supported by a gift from Google.

The authors wish to recognize and acknowledge the very significant cultural role and reverence that the summit of Maunakea has always had within the Native Hawaiian community. We are most fortunate to have the opportunity to conduct observations from this mountain. We appreciate the expert assistance of the staff at the various observatories where data were obtained.  

\end{acknowledgments}

\facilities{Lick(APF \& Shane), Gemini Gillett (Maroon-X \& IGRINS-2)}
\software{Astropy \citep{astropy:2013, astropy:2018, astropy:2022},
          Matplotlib \citep{Hunter:2007},
          Numpy \citep{harris2020array},
          Pandas \citep{mckinney-proc-scipy-2010, reback2020pandas},
          PyAstronomy \citep{pya},
          Scipy \citep{2020SciPy-NMeth}}

\appendix
\section{Telluric Correction} \label{sec:6.1}

\begin{figure*}
    \centering
    \includegraphics[scale=0.4]{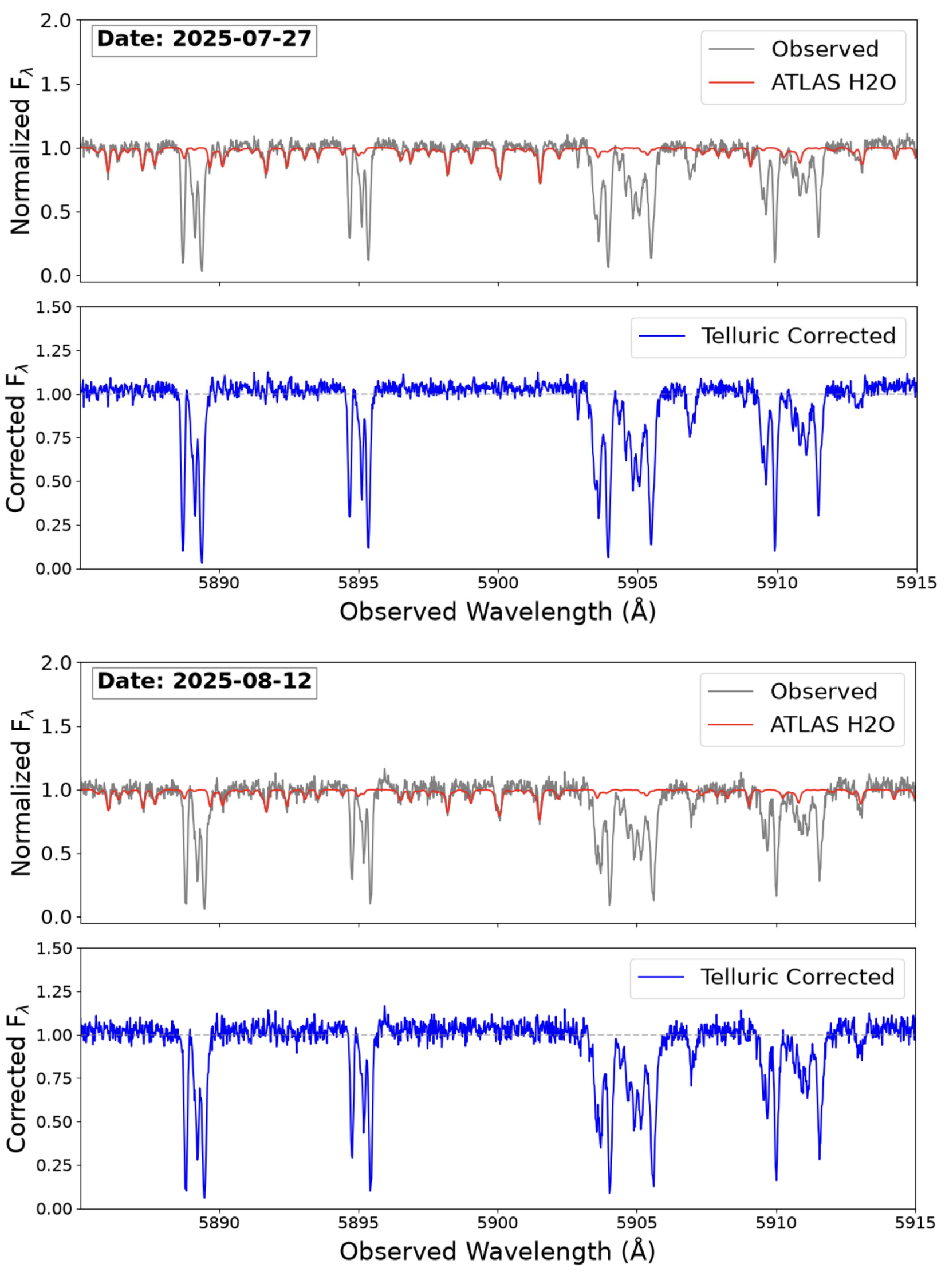}
    \caption{A few representative examples of APF spectra (gray) around \NaID absorption lines corrected for telluric bands using the model H$_{2}$O spectra (red) within our high-resolution spectral time series. The final telluric-corrected spectra are shown in the bottom panel for each epoch (blue).}
    \label{fig:Telluric_correction}
\end{figure*}

Telluric correction is applied to the APF and M-X spectra using a two-step empirical fit against a synthetic atlas of H$_{2}$O (around \NaID) and O$_{2}$ (around \KI) telluric absorption bands. We present a few representative examples of telluric corrections in Figure \ref{fig:Telluric_correction}.

Both the normalized data spectrum and the template are converted to optical-depth scales ($\tau = -\ln F$), mean-subtracted, and cross-correlated. A parabolic interpolation around the peak gives a subpixel wavelength shift $\delta\lambda$, which is applied to the template before further fitting. Pixels coinciding with known SN absorption features are masked during this step.

Because the effective H$_{2}$O and O$_{2}$ columns vary with nightly airmass and precipitable water vapor, a pixel-by-pixel scaling exponent $\alpha(\lambda)$ is derived from the ratio $\ln F_\mathrm{data} / \ln F_\mathrm{telluric}$ over moderately absorbing telluric pixels ($0.01 < F_\mathrm{telluric} < 0.95$). A second-degree polynomial is fit to $\alpha(\lambda)$ after $3\sigma$ outlier rejection, giving a wavelength-dependent correction $T(\lambda) = F_\mathrm{telluric}(\lambda - \delta\lambda)^{\alpha(\lambda)}$. The normalized spectrum is then divided by $T(\lambda)$, and pixels where $T < 0.05$ (saturated telluric cores) are masked.

\section{Voigt-Profile Fitting Components} \label{sec:6.2}

Tables \ref{tab:Voigt_NaID} and \ref{tab:Voigt_CaII} show Voigt-profile results from multi-components to \NaID (D$_{1}$ + D$_{2}$) Ca\,II (H \& K). 

\begin{deluxetable}{lccc}
\tablecaption{Voigt Profile Fits: \NaID Absorption Column Densities}
\tablehead{
\colhead{Component$^{*}$} & \colhead{Velocity$^{\dag}$} & \colhead{Column density} & \colhead{$b$} \\
\colhead{} & \colhead{(km s$^{-1}$)} & \colhead{\ensuremath{\log(N/\mathrm{cm}^{-2})}} & \colhead{(km s$^{-1}$)}
}
\startdata
     1 & $-$140.44 $\pm$ 1.53 & 10.97 $\pm$ 0.33 & 0.71$^{\ddag}$ \\
     2 & $-$120.25 $\pm$ 2.86 & 10.90 $\pm$ 0.39 & 1.71$^{\ddag}$ \\
     3 & $-$110.30 $\pm$ 1.76 & 11.70 $\pm$ 0.49 & 3.45 $\pm$ 2.24 \\
     4 & $-$102.09 $\pm$ 1.03 & 12.00 $\pm$ 0.36 & 4.29 $\pm$ 2.02 \\
     5 & $-$85.38  $\pm$ 0.37 & 12.37 $\pm$ 0.71 & 3.72 $\pm$ 1.04 \\
     6 & $-$65.55  $\pm$ 2.92 & 11.20 $\pm$ 0.28 & 4.33 $\pm$ 4.23 \\
     7 & $-$52.60  $\pm$ 1.50 & 11.57 $\pm$ 0.35 & 3.67 $\pm$ 2.50 \\
     8 & $-$41.02  $\pm$ 1.61 & 11.73 $\pm$ 0.23 & 3.90 $\pm$ 2.63 \\
     9 & $-$28.70  $\pm$ 1.98 & 11.99 $\pm$ 0.12 & 7.80 $\pm$ 3.40 \\
     10 & $-$6.73  $\pm$ 0.55 & 12.27 $\pm$ 0.06 & 5.67 $\pm$ 1.05 \\
     11 & +63.31 $\pm$ 2.31 & 11.42 $\pm$ 0.17 & 5.35 $\pm$ 3.52 \\
     12 & +71.68 $\pm$ 5.29 & 10.95 $\pm$ 2.04 & 2.36$^{\ddag}$
\enddata
\tablenotetext{*}{Components marked in Figure \ref{fig:NaID}.}
\tablenotetext{\dag}{With respect to the rest frame of NGC~7331.}
\tablenotetext{\ddag}{Widths associated with weaker absorption lines are poorly constrained and should be considered limits $\lesssim$3 km s$^{-1}$, comparable to the the resolution of M-X around \NaID wavelengths.}
\label{tab:Voigt_NaID}
\end{deluxetable}

\begin{deluxetable}{lccc}
\tablecaption{Voigt Profile Fits: Ca\,II (H \& K) Absorption Column Densities}
\tablehead{
\colhead{Component} & \colhead{Velocity$^{\dag}$} & \colhead{Column density} & \colhead{$b$} \\
\colhead{} & \colhead{(km s$^{-1}$)} & \colhead{\ensuremath{\log(N/\mathrm{cm}^{-2})}} & \colhead{(km s$^{-1}$)}
}
\startdata
     1 & $-$114.84 $\pm$ 1.39 & 11.52 $\pm$ 0.19 & 7.73 $\pm$ 4.67 \\
     2 & $-$107.07 $\pm$ 1.66 & 11.46 $\pm$ 0.25 & 8.29 $\pm$ 1.78 \\
     3 & $-$97.08 $\pm$ 0.69 & 11.09 $\pm$ 0.14 & 2.52 $\pm$ 1.18 \\
     4 & $-$83.36 $\pm$ 0.32 & 12.01 $\pm$ 0.03 & 7.74 $\pm$ 0.54 \\
     5 & $-$64.20 $\pm$ 0.79 & 11.51 $\pm$ 0.09 & 5.37 $\pm$ 1.87 \\
     6 & $-$52.99 $\pm$ 0.68 & 11.86 $\pm$ 0.18 & 5.03 $\pm$ 1.52 \\
     7 & $-$48.58 $\pm$ 0.72 & 10.25 $\pm$ 0.67 & 9.87 $\pm$ 0.67 \\
     8 & $-$33.90 $\pm$ 0.92 & 12.30 $\pm$ 0.05 & 13.22 $\pm$ 1.59 \\
     9 & $-$6.59 $\pm$ 0.47 & 12.17 $\pm$ 0.04 & 9.85 $\pm$ 1.01 \\
     10 & +8.18 $\pm$ 0.74 & 10.90 $\pm$ 0.19 & 0.50 \ddag \\
     11 & +21.23 $\pm$ 0.76 & 11.97 $\pm$ 0.09 & 11.65 $\pm$ 2.19 \\
     12 & +38.96 $\pm$ 1.67 & 11.32 $\pm$ 0.21 & 7.49 $\pm$ 2.27 \\
     13 & +66.16 $\pm$ 2.22 & 11.69 $\pm$ 0.08 & 17.83 $\pm$ 3.74
\enddata
\tablenotetext{\dag}{With respect to the rest frame of NGC~7331.}
\tablenotetext{\ddag}{Widths associated with weaker absorption lines are poorly constrained and should be considered limits $\lesssim$ 3 km s$^{-1}$, comparable to the the resolution of APF around Ca\,II (H \& K) wavelengths.}
\label{tab:Voigt_CaII}
\end{deluxetable}

\bibliography{sample701}

@ARTICLE{deBlok08,
       author = {{de Blok}, W.~J.~G. and {Walter}, F. and {Brinks}, E. and {Trachternach}, C. and {Oh}, S.-H. and {Kennicutt}, Jr., R.~C.},
        title = "{High-Resolution Rotation Curves and Galaxy Mass Models from THINGS}",
      journal = {\aj},
         year = 2008,
        month = dec,
       volume = {136},
       number = {6},
        pages = {2648-2719},
          doi = {10.1088/0004-6256/136/6/2648},
archivePrefix = {arXiv},
       eprint = {0810.2100},
 primaryClass = {astro-ph},
       adsurl = {https://ui.adsabs.harvard.edu/abs/2008AJ....136.2648D}
}

@ARTICLE{Sternberg11,
       author = {{Sternberg}, A. and {Gal-Yam}, A. and {Simon}, J.~D. and {Leonard}, D.~C. and {Quimby}, R.~M. and {Phillips}, M.~M. and {Morrell}, N. and {Thompson}, I.~B. and {Ivans}, I. and {Marshall}, J.~L. and {Filippenko}, A.~V. and {Marcy}, G.~W. and {Bloom}, J.~S. and {Patat}, F. and {Foley}, R.~J. and {Yong}, D. and {Penprase}, B.~E. and {Beeler}, D.~J. and {Allende Prieto}, C. and {Stringfellow}, G.~S.},
        title = "{Circumstellar Material in Type Ia Supernovae via Sodium Absorption Features}",
      journal = {Science},
         year = 2011,
        month = aug,
       volume = {333},
       number = {6044},
        pages = {856},
          doi = {10.1126/science.1203836},
archivePrefix = {arXiv},
       eprint = {1108.3664},
 primaryClass = {astro-ph.HE},
       adsurl = {https://ui.adsabs.harvard.edu/abs/2011Sci...333..856S}
}

@ARTICLE{Phillips13,
       author = {{Phillips}, M.~M. and {Simon}, Joshua D. and {Morrell}, Nidia and {Burns}, Christopher R. and {Cox}, Nick L.~J. and {Foley}, Ryan J. and {Karakas}, Amanda I. and {Patat}, F. and {Sternberg}, A. and {Williams}, R.~E. and {Gal-Yam}, A. and {Hsiao}, E.~Y. and {Leonard}, D.~C. and {Persson}, Sven E. and {Stritzinger}, Maximilian and {Thompson}, I.~B. and {Campillay}, Abdo and {Contreras}, Carlos and {Folatelli}, Gast{\'o}n and {Freedman}, Wendy L. and {Hamuy}, Mario and {Roth}, Miguel and {Shields}, Gregory A. and {Suntzeff}, Nicholas B. and {Chomiuk}, Laura and {Ivans}, Inese I. and {Madore}, Barry F. and {Penprase}, B.~E. and {Perley}, Daniel and {Pignata}, G. and {Preston}, G. and {Soderberg}, Alicia M.},
        title = "{On the Source of the Dust Extinction in Type Ia Supernovae and the Discovery of Anomalously Strong Na I Absorption}",
      journal = {\apj},
         year = 2013,
        month = dec,
       volume = {779},
       number = {1},
          eid = {38},
        pages = {38},
          doi = {10.1088/0004-637X/779/1/38},
archivePrefix = {arXiv},
       eprint = {1311.0147},
 primaryClass = {astro-ph.CO},
       adsurl = {https://ui.adsabs.harvard.edu/abs/2013ApJ...779...38P}
}

@ARTICLE{Prada96,
       author = {{Prada}, F. and {Gutierrez}, C.~M. and {Peletier}, R.~F. and {McKeith}, C.~D.},
        title = "{A Counterrotating Bulge in the S(b) Galaxy NGC 7331}",
      journal = {\apjl},
         year = 1996,
        month = may,
       volume = {463},
        pages = {L9},
          doi = {10.1086/310044},
archivePrefix = {arXiv},
       eprint = {astro-ph/9602142},
 primaryClass = {astro-ph},
       adsurl = {https://ui.adsabs.harvard.edu/abs/1996ApJ...463L...9P}
}

@ARTICLE{Maguire13,
       author = {{Maguire}, K. and {Sullivan}, M. and {Patat}, F. and {Gal-Yam}, A. and {Hook}, I.~M. and {Dhawan}, S. and {Howell}, D.~A. and {Mazzali}, P. and {Nugent}, P.~E. and {Pan}, Y.-C. and {Podsiadlowski}, P. and {Simon}, J.~D. and {Sternberg}, A. and {Valenti}, S. and {Baltay}, C. and {Bersier}, D. and {Blagorodnova}, N. and {Chen}, T.-W. and {Ellman}, N. and {Feindt}, U. and {F{\"o}rster}, F. and {Fraser}, M. and {Gonz{\'a}lez-Gait{\'a}n}, S. and {Graham}, M.~L. and {Guti{\'e}rrez}, C. and {Hachinger}, S. and {Hadjiyska}, E. and {Inserra}, C. and {Knapic}, C. and {Laher}, R.~R. and {Leloudas}, G. and {Margheim}, S. and {McKinnon}, R. and {Molinaro}, M. and {Morrell}, N. and {Ofek}, E.~O. and {Rabinowitz}, D. and {Rest}, A. and {Sand}, D. and {Smareglia}, R. and {Smartt}, S.~J. and {Taddia}, F. and {Walker}, E.~S. and {Walton}, N.~A. and {Young}, D.~R.},
        title = "{A statistical analysis of circumstellar material in Type Ia supernovae}",
      journal = {\mnras},
         year = 2013,
        month = nov,
       volume = {436},
       number = {1},
        pages = {222-240},
          doi = {10.1093/mnras/stt1586},
archivePrefix = {arXiv},
       eprint = {1308.3899},
 primaryClass = {astro-ph.SR},
       adsurl = {https://ui.adsabs.harvard.edu/abs/2013MNRAS.436..222M}
}

@ARTICLE{Ruiter25,
       author = {{Ruiter}, Ashley Jade and {Seitenzahl}, Ivo Rolf},
        title = "{Type Ia supernova progenitors: a contemporary view of a long-standing puzzle}",
      journal = {AAPR},
         year = 2025,
        month = dec,
       volume = {33},
       number = {1},
          eid = {1},
        pages = {1},
          doi = {10.1007/s00159-024-00158-9},
archivePrefix = {arXiv},
       eprint = {2412.01766},
 primaryClass = {astro-ph.SR},
       adsurl = {https://ui.adsabs.harvard.edu/abs/2025A&ARv..33....1R}
}

@ARTICLE{Liu23,
       author = {{Liu}, Zheng-Wei and {R{\"o}pke}, Friedrich K. and {Han}, Zhanwen},
        title = "{Type Ia Supernova Explosions in Binary Systems: A Review}",
      journal = {Research in Astronomy and Astrophysics},
         year = 2023,
        month = aug,
       volume = {23},
       number = {8},
          eid = {082001},
        pages = {082001},
          doi = {10.1088/1674-4527/acd89e},
archivePrefix = {arXiv},
       eprint = {2305.13305},
 primaryClass = {astro-ph.HE},
       adsurl = {https://ui.adsabs.harvard.edu/abs/2023RAA....23h2001L}
}

@ARTICLE{Jha19,
       author = {{Jha}, Saurabh W. and {Maguire}, Kate and {Sullivan}, Mark},
        title = "{Observational properties of thermonuclear supernovae}",
      journal = {Nature Astronomy},
         year = 2019,
        month = aug,
       volume = {3},
        pages = {706-716},
          doi = {10.1038/s41550-019-0858-0},
archivePrefix = {arXiv},
       eprint = {1908.02303},
 primaryClass = {astro-ph.HE},
       adsurl = {https://ui.adsabs.harvard.edu/abs/2019NatAs...3..706J}
}

@INCOLLECTION{Taubenberger17,
       author = {{Taubenberger}, Stefan},
        title = "{The Extremes of Thermonuclear Supernovae}",
    booktitle = {Handbook of Supernovae},
         year = 2017,
       editor = {{Alsabti}, Athem W. and {Murdin}, Paul},
        pages = {317},
          doi = {10.1007/978-3-319-21846-5_37},
       adsurl = {https://ui.adsabs.harvard.edu/abs/2017hsn..book..317T}
}

@ARTICLE{Whelan73,
       author = {{Whelan}, John and {Iben}, Jr., Icko},
        title = "{Binaries and Supernovae of Type I}",
      journal = {ApJ},
         year = 1973,
        month = dec,
       volume = {186},
        pages = {1007-1014},
          doi = {10.1086/152565},
       adsurl = {https://ui.adsabs.harvard.edu/abs/1973ApJ...186.1007W}
}

@article{Webbink84,
  title = {Double White Dwarfs as Progenitors of {{R Coronae Borealis}} Stars and {{Type I}} Supernovae},
  author = {Webbink, R. F.},
  year = {1984},
  month = feb,
  journal = {ApJ},
  volume = {277},
  pages = {355--360},
  issn = {0004-637X},
  doi = {10.1086/161701},
  urldate = {2018-01-17},
  langid = {english}
}

@article{Iben84,
  title = {Supernovae of Type {{I}} as End Products of the Evolution of Binaries with Components of Moderate Initial Mass ({{M}} Not Greater than about 9 Solar Masses)},
  author = {Iben, I. and Tutukov, A. V.},
  year = {1984},
  month = feb,
  journal = {ApJS},
  volume = {54},
  pages = {335--372},
  issn = {0067-0049},
  doi = {10.1086/190932},
  urldate = {2018-01-17},
  langid = {english}
}

@article{Katz12,
  title = {The Rate of {{WD-WD}} Head-on Collisions May Be as High as the {{SNe Ia}} Rate},
  author = {Katz, Boaz and Dong, Subo},
  year = {2012},
  month = nov,
  eprint = {1211.4584},
  urldate = {2022-04-15},
  archiveprefix = {arXiv}
}

@article{Kushnir13,
  title = {Head-on {{Collisions}} of {{White Dwarfs}} in {{Triple Systems Could Explain Type Ia Supernovae}}},
  author = {Kushnir, D. and Katz, B. and Dong, S. and Livne, E. and Fern{\'a}ndez, R.},
  year = {2013},
  month = dec,
  journal = {ApJL},
  volume = {778},
  eprint = {1303.1180},
  pages = {L37},
  doi = {10.1088/2041-8205/778/2/L37},
  archiveprefix = {arXiv}
}

@ARTICLE{Zenati19b,
       author = {{Zenati}, Yossef and {Perets}, Hagai B. and {Toonen}, Silvia},
        title = "{Neutron star-white dwarf mergers: early evolution, physical properties, and outcomes}",
      journal = {\mnras},
         year = 2019,
        month = jun,
       volume = {486},
       number = {2},
        pages = {1805-1813},
          doi = {10.1093/mnras/stz316},
archivePrefix = {arXiv},
       eprint = {1807.09777},
 primaryClass = {astro-ph.HE},
       adsurl = {https://ui.adsabs.harvard.edu/abs/2019MNRAS.486.1805Z}
}

@ARTICLE{Raskin_Kasen13,
       author = {{Raskin}, Cody and {Kasen}, Daniel},
        title = "{Tidal Tail Ejection as a Signature of Type Ia Supernovae from White Dwarf Mergers}",
      journal = {\apj},
         year = 2013,
        month = jul,
       volume = {772},
       number = {1},
          eid = {1},
        pages = {1},
          doi = {10.1088/0004-637X/772/1/1},
archivePrefix = {arXiv},
       eprint = {1304.4957},
 primaryClass = {astro-ph.HE},
       adsurl = {https://ui.adsabs.harvard.edu/abs/2013ApJ...772....1R}
}

@ARTICLE{Shen12,
       author = {{Shen}, Ken J. and {Bildsten}, Lars and {Kasen}, Daniel and {Quataert}, Eliot},
        title = "{The Long-term Evolution of Double White Dwarf Mergers}",
      journal = {\apj},
         year = 2012,
        month = mar,
       volume = {748},
       number = {1},
          eid = {35},
        pages = {35},
          doi = {10.1088/0004-637X/748/1/35},
archivePrefix = {arXiv},
       eprint = {1108.4036},
 primaryClass = {astro-ph.HE},
       adsurl = {https://ui.adsabs.harvard.edu/abs/2012ApJ...748...35S}
}

@ARTICLE{Schwab16,
       author = {{Schwab}, Josiah and {Quataert}, Eliot and {Kasen}, Daniel},
        title = "{The evolution and fate of super-Chandrasekhar mass white dwarf merger remnants}",
      journal = {\mnras},
         year = 2016,
        month = dec,
       volume = {463},
       number = {4},
        pages = {3461-3475},
          doi = {10.1093/mnras/stw2249},
archivePrefix = {arXiv},
       eprint = {1606.02300},
 primaryClass = {astro-ph.SR},
       adsurl = {https://ui.adsabs.harvard.edu/abs/2016MNRAS.463.3461S}
}

@ARTICLE{Patat07,
       author = {{Patat}, F. and {Chandra}, P. and {Chevalier}, R. and {Justham}, S. and {Podsiadlowski}, Ph. and {Wolf}, C. and {Gal-Yam}, A. and {Pasquini}, L. and {Crawford}, I.~A. and {Mazzali}, P.~A. and {Pauldrach}, A.~W.~A. and {Nomoto}, K. and {Benetti}, S. and {Cappellaro}, E. and {Elias-Rosa}, N. and {Hillebrandt}, W. and {Leonard}, D.~C. and {Pastorello}, A. and {Renzini}, A. and {Sabbadin}, F. and {Simon}, J.~D. and {Turatto}, M.},
        title = "{Detection of Circumstellar Material in a Normal Type Ia Supernova}",
      journal = {Science},
         year = 2007,
        month = aug,
       volume = {317},
       number = {5840},
        pages = {924},
          doi = {10.1126/science.1143005},
archivePrefix = {arXiv},
       eprint = {0707.2793},
 primaryClass = {astro-ph},
       adsurl = {https://ui.adsabs.harvard.edu/abs/2007Sci...317..924P}
}

@ARTICLE{Blondin09,
       author = {{Blondin}, S. and {Prieto}, J.~L. and {Patat}, F. and {Challis}, P. and {Hicken}, M. and {Kirshner}, R.~P. and {Matheson}, T. and {Modjaz}, M.},
        title = "{A Second Case of Variable Na I D Lines in a Highly Reddened Type Ia Supernova}",
      journal = {\apj},
         year = 2009,
        month = mar,
       volume = {693},
       number = {1},
        pages = {207-215},
          doi = {10.1088/0004-637X/693/1/207},
archivePrefix = {arXiv},
       eprint = {0811.0002},
 primaryClass = {astro-ph},
       adsurl = {https://ui.adsabs.harvard.edu/abs/2009ApJ...693..207B}
}

@ARTICLE{Simon09,
       author = {{Simon}, Joshua D. and {Gal-Yam}, Avishay and {Gnat}, Orly and {Quimby}, Robert M. and {Ganeshalingam}, Mohan and {Silverman}, Jeffrey M. and {Blondin}, Stephane and {Li}, Weidong and {Filippenko}, Alexei V. and {Wheeler}, J. Craig and {Kirshner}, Robert P. and {Patat}, Ferdinando and {Nugent}, Peter and {Foley}, Ryan J. and {Vogt}, Steven S. and {Butler}, R. Paul and {Peek}, Kathryn M.~G. and {Rosolowsky}, Erik and {Herczeg}, Gregory J. and {Sauer}, Daniel N. and {Mazzali}, Paolo A.},
        title = "{Variable Sodium Absorption in a Low-extinction Type Ia Supernova}",
      journal = {\apj},
         year = 2009,
        month = sep,
       volume = {702},
       number = {2},
        pages = {1157-1170},
          doi = {10.1088/0004-637X/702/2/1157},
archivePrefix = {arXiv},
       eprint = {0907.1083},
 primaryClass = {astro-ph.HE},
       adsurl = {https://ui.adsabs.harvard.edu/abs/2009ApJ...702.1157S}
}

@ARTICLE{Heger22,
       author = {{Heger}, Mary Lea},
        title = "{Further study of the sodium lines in class B stars}",
      journal = {Lick Observatory Bulletin},
         year = 1922,
        month = jan,
       volume = {10},
       number = {337},
        pages = {141-145},
          doi = {10.5479/ADS/bib/1922LicOB.10.141H},
       adsurl = {https://ui.adsabs.harvard.edu/abs/1922LicOB..10..141H}
}

@ARTICLE{Merrill34,
       author = {{Merrill}, P.~W.},
        title = "{Unidentified Interstellar Lines}",
      journal = {\pasp},
         year = 1934,
        month = aug,
       volume = {46},
       number = {272},
        pages = {206-207},
          doi = {10.1086/124460},
       adsurl = {https://ui.adsabs.harvard.edu/abs/1934PASP...46..206M}
}

@ARTICLE{Vogt14,
       author = {{Vogt}, Steven S. and {Radovan}, Matthew and {Kibrick}, Robert and {Butler}, R. Paul and {Alcott}, Barry and {Allen}, Steve and {Arriagada}, Pamela and {Bolte}, Mike and {Burt}, Jennifer and {Cabak}, Jerry and {Chloros}, Kostas and {Cowley}, David and {Deich}, William and {Dupraw}, Brian and {Earthman}, Wayne and {Epps}, Harland and {Faber}, Sandra and {Fischer}, Debra and {Gates}, Elinor and {Hilyard}, David and {Holden}, Brad and {Johnston}, Ken and {Keiser}, Sandy and {Kanto}, Dick and {Katsuki}, Myra and {Laiterman}, Lee and {Lanclos}, Kyle and {Laughlin}, Greg and {Lewis}, Jeff and {Lockwood}, Chris and {Lynam}, Paul and {Marcy}, Geoffrey and {McLean}, Maureen and {Miller}, Joe and {Misch}, Tony and {Peck}, Michael and {Pfister}, Terry and {Phillips}, Andrew and {Rivera}, Eugenio and {Sandford}, Dale and {Saylor}, Mike and {Stover}, Richard and {Thompson}, Matthew and {Walp}, Bernie and {Ward}, James and {Wareham}, John and {Wei}, Mingzhi and {Wright}, Chris},
        title = "{APF{\textemdash}The Lick Observatory Automated Planet Finder}",
      journal = {\pasp},
         year = 2014,
        month = apr,
       volume = {126},
       number = {938},
        pages = {359},
          doi = {10.1086/676120},
archivePrefix = {arXiv},
       eprint = {1402.6684},
 primaryClass = {astro-ph.IM},
       adsurl = {https://ui.adsabs.harvard.edu/abs/2014PASP..126..359V}
}

@INPROCEEDINGS{Seifahrt18,
       author = {{Seifahrt}, Andreas and {St{\"u}rmer}, Julian and {Bean}, Jacob L. and {Schwab}, Christian},
        title = "{MAROON-X: a radial velocity spectrograph for the Gemini Observatory}",
    booktitle = {Ground-based and Airborne Instrumentation for Astronomy VII},
         year = 2018,
       editor = {{Evans}, Christopher J. and {Simard}, Luc and {Takami}, Hideki},
       series = {Society of Photo-Optical Instrumentation Engineers (SPIE) Conference Series},
       volume = {10702},
        month = jul,
          eid = {107026D},
        pages = {107026D},
          doi = {10.1117/12.2312936},
archivePrefix = {arXiv},
       eprint = {1805.09276},
 primaryClass = {astro-ph.IM},
       adsurl = {https://ui.adsabs.harvard.edu/abs/2018SPIE10702E..6DS}
}

@ARTICLE{Suh25,
       author = {{Suh}, H.},
        title = "{IGRINS-2: High-Resolution Near-Infrared Spectrograph at Gemini North}",
      journal = {The NOIRLab Mirror},
         year = 2025,
        month = jan,
       volume = {8},
        pages = {14},
       adsurl = {https://ui.adsabs.harvard.edu/abs/2025Mirro...8...14S}
}

@ARTICLE{Maoz14,
       author = {{Maoz}, Dan and {Mannucci}, Filippo and {Nelemans}, Gijs},
        title = "{Observational Clues to the Progenitors of Type Ia Supernovae}",
      journal = {\araa},
         year = 2014,
        month = aug,
       volume = {52},
        pages = {107-170},
          doi = {10.1146/annurev-astro-082812-141031},
archivePrefix = {arXiv},
       eprint = {1312.0628},
 primaryClass = {astro-ph.CO},
       adsurl = {https://ui.adsabs.harvard.edu/abs/2014ARA&A..52..107M}
}

@ARTICLE{Dilday12,
       author = {{Dilday}, B. and {Howell}, D.~A. and {Cenko}, S.~B. and {Silverman}, J.~M. and {Nugent}, P.~E. and {Sullivan}, M. and {Ben-Ami}, S. and {Bildsten}, L. and {Bolte}, M. and {Endl}, M. and {Filippenko}, A.~V. and {Gnat}, O. and {Horesh}, A. and {Hsiao}, E. and {Kasliwal}, M.~M. and {Kirkman}, D. and {Maguire}, K. and {Marcy}, G.~W. and {Moore}, K. and {Pan}, Y. and {Parrent}, J.~T. and {Podsiadlowski}, P. and {Quimby}, R.~M. and {Sternberg}, A. and {Suzuki}, N. and {Tytler}, D.~R. and {Xu}, D. and {Bloom}, J.~S. and {Gal-Yam}, A. and {Hook}, I.~M. and {Kulkarni}, S.~R. and {Law}, N.~M. and {Ofek}, E.~O. and {Polishook}, D. and {Poznanski}, D.},
        title = "{PTF 11kx: A Type Ia Supernova with a Symbiotic Nova Progenitor}",
      journal = {Science},
         year = 2012,
        month = aug,
       volume = {337},
       number = {6097},
        pages = {942},
          doi = {10.1126/science.1219164},
archivePrefix = {arXiv},
       eprint = {1207.1306},
 primaryClass = {astro-ph.CO},
       adsurl = {https://ui.adsabs.harvard.edu/abs/2012Sci...337..942D}
}

@ARTICLE{ONiell25,
       author = {{O'Neill}, D. and {Ackley}, K. and {Dyer}, M. and {Lyman}, J. and {Ulaczyk}, K. and {Steeghs}, D. and {Galloway}, D. and {Dhillon}, V. and {O'Brien}, P. and {Ramsay}, G. and {Noysena}, K. and {Kotak}, R. and {Breton}, R. and {Casares}, J. and {Nuttall}, L. and {Godson}, B. and {Killestein}, T. and {Kumar}, A. and {Pursiainen}, M.},
        title = "{GOTO Transient Discovery Report for 2025-07-14}",
      journal = {Transient Name Server Discovery Report},
         year = 2025,
        month = jul,
       volume = {2025-2688},
        pages = {1},
       adsurl = {https://ui.adsabs.harvard.edu/abs/2025TNSTR2688....1O}
}

@ARTICLE{Andrews25,
       author = {{Andrews}, M. and {Farah}, J. and {Wynn}, K. and {Bosteroem}, A. and {Howell}, D.~A. and {McCully}, C.},
        title = "{Global SN Project Transient Classification Report for 2025-07-14}",
      journal = {Transient Name Server Classification Report},
         year = 2025,
        month = jul,
       volume = {2025-2699},
        pages = {1},
       adsurl = {https://ui.adsabs.harvard.edu/abs/2025TNSCR2699....1A}
}

@INPROCEEDINGS{Dyer22,
       author = {{Dyer}, Martin J. and {Ackley}, Kendall and {Lyman}, Joe and {Ulaczyk}, Krzysztof and {Steeghs}, Danny and {Galloway}, Duncan K. and {Dhillon}, Vik S. and {O'Brien}, Paul and {Ramsay}, Gavin and {Noysena}, Kanthanakorn and {Kotak}, Rubina and {Breton}, Rene and {Nuttall}, Laura and {Pall{\'e}}, Enric and {Pollacco}, Don},
        title = "{The Gravitational-wave Optical Transient Observer (GOTO)}",
    booktitle = {Ground-based and Airborne Telescopes IX},
         year = 2022,
       editor = {{Marshall}, Heather K. and {Spyromilio}, Jason and {Usuda}, Tomonori},
       series = {Society of Photo-Optical Instrumentation Engineers (SPIE) Conference Series},
       volume = {12182},
        month = aug,
          eid = {121821Y},
        pages = {121821Y},
          doi = {10.1117/12.2629369},
archivePrefix = {arXiv},
       eprint = {2208.14901},
 primaryClass = {astro-ph.IM},
       adsurl = {https://ui.adsabs.harvard.edu/abs/2022SPIE12182E..1YD}
}

@ARTICLE{Freedman01,
       author = {{Freedman}, Wendy L. and {Madore}, Barry F. and {Gibson}, Brad K. and {Ferrarese}, Laura and {Kelson}, Daniel D. and {Sakai}, Shoko and {Mould}, Jeremy R. and {Kennicutt}, Jr., Robert C. and {Ford}, Holland C. and {Graham}, John A. and {Huchra}, John P. and {Hughes}, Shaun M.~G. and {Illingworth}, Garth D. and {Macri}, Lucas M. and {Stetson}, Peter B.},
        title = "{Final Results from the Hubble Space Telescope Key Project to Measure the Hubble Constant}",
      journal = {\apj},
         year = 2001,
        month = may,
       volume = {553},
       number = {1},
        pages = {47-72},
          doi = {10.1086/320638},
archivePrefix = {arXiv},
       eprint = {astro-ph/0012376},
 primaryClass = {astro-ph},
       adsurl = {https://ui.adsabs.harvard.edu/abs/2001ApJ...553...47F}
}

@ARTICLE{Patra18,
       author = {{Patra}, Narendra Nath},
        title = "{Molecular scale height in NGC 7331}",
      journal = {\mnras},
         year = 2018,
        month = aug,
       volume = {478},
       number = {4},
        pages = {4931-4938},
          doi = {10.1093/mnras/sty1512},
       adsurl = {https://ui.adsabs.harvard.edu/abs/2018MNRAS.478.4931P}
}

@ARTICLE{Tosaki97,
       author = {{Tosaki}, T. and {Shioya}, Y.},
        title = "{Molecular Gas in the Poststarburst Galaxy NGC 7331}",
      journal = {\apj},
         year = 1997,
        month = jul,
       volume = {484},
       number = {2},
        pages = {664-671},
          doi = {10.1086/304361},
       adsurl = {https://ui.adsabs.harvard.edu/abs/1997ApJ...484..664T}
}

@ARTICLE{Hobbs74,
       author = {{Hobbs}, L.~M.},
        title = "{A comparison of interstellar Na I, Ca II, and K I absorption.}",
      journal = {\apj},
         year = 1974,
        month = jul,
       volume = {191},
        pages = {381-393},
          doi = {10.1086/152976},
       adsurl = {https://ui.adsabs.harvard.edu/abs/1974ApJ...191..381H}
}

@ARTICLE{Poznanski12,
       author = {{Poznanski}, Dovi and {Prochaska}, J. Xavier and {Bloom}, Joshua S.},
        title = "{An empirical relation between sodium absorption and dust extinction}",
      journal = {\mnras},
         year = 2012,
        month = oct,
       volume = {426},
       number = {2},
        pages = {1465-1474},
          doi = {10.1111/j.1365-2966.2012.21796.x},
archivePrefix = {arXiv},
       eprint = {1206.6107},
 primaryClass = {astro-ph.IM},
       adsurl = {https://ui.adsabs.harvard.edu/abs/2012MNRAS.426.1465P}
}

@ARTICLE{Graham15,
       author = {{Graham}, M.~L. and {Valenti}, S. and {Fulton}, B.~J. and {Weiss}, L.~M. and {Shen}, K.~J. and {Kelly}, P.~L. and {Zheng}, W. and {Filippenko}, A.~V. and {Marcy}, G.~W. and {Howell}, D.~A. and {Burt}, J. and {Rivera}, E.~J.},
        title = "{Time-Varying Potassium in High-Resolution Spectra of the Type Ia Supernova 2014j}",
      journal = {\apj},
         year = 2015,
        month = mar,
       volume = {801},
       number = {2},
          eid = {136},
        pages = {136},
          doi = {10.1088/0004-637X/801/2/136},
archivePrefix = {arXiv},
       eprint = {1412.0653},
 primaryClass = {astro-ph.SR},
       adsurl = {https://ui.adsabs.harvard.edu/abs/2015ApJ...801..136G}
}

@ARTICLE{Baldwin85,
       author = {{Baldwin}, J.~A. and {Phillips}, M.~M. and {Carswell}, R.~F.},
        title = "{Do some QSO low-ionization absorption systems ARISE in galactic discs ?}",
      journal = {\mnras},
         year = 1985,
        month = sep,
       volume = {216},
        pages = {41P-44},
          doi = {10.1093/mnras/216.1.41P},
       adsurl = {https://ui.adsabs.harvard.edu/abs/1985MNRAS.216P..41B}
}

@ARTICLE{Welty_Hobbs_Kulkarni94,
       author = {{Welty}, Daniel E. and {Hobbs}, L.~M. and {Kulkarni}, Varsha P.},
        title = "{A High-Resolution Survey of Interstellar NA i D 1 Lines}",
      journal = {\apj},
         year = 1994,
        month = nov,
       volume = {436},
        pages = {152},
          doi = {10.1086/174889},
       adsurl = {https://ui.adsabs.harvard.edu/abs/1994ApJ...436..152W}
}

@ARTICLE{Morton03,
       author = {{Morton}, Donald C.},
        title = "{Atomic Data for Resonance Absorption Lines. III. Wavelengths Longward of the Lyman Limit for the Elements Hydrogen to Gallium}",
      journal = {\apjs},
         year = 2003,
        month = nov,
       volume = {149},
       number = {1},
        pages = {205-238},
          doi = {10.1086/377639},
       adsurl = {https://ui.adsabs.harvard.edu/abs/2003ApJS..149..205M}
}

@INPROCEEDINGS{Asplund05,
       author = {{Asplund}, M. and {Grevesse}, N. and {Sauval}, A.~J.},
        title = "{The Solar Chemical Composition}",
    booktitle = {Cosmic Abundances as Records of Stellar Evolution and Nucleosynthesis},
         year = 2005,
       editor = {{Barnes}, III, Thomas G. and {Bash}, Frank N.},
       series = {Astronomical Society of the Pacific Conference Series},
       volume = {336},
        month = sep,
        pages = {25},
       adsurl = {https://ui.adsabs.harvard.edu/abs/2005ASPC..336...25A}
}

@ARTICLE{Hamuy03,
       author = {{Hamuy}, Mario and {Phillips}, M.~M. and {Suntzeff}, Nicholas B. and {Maza}, Jos{\'e} and {Gonz{\'a}lez}, L.~E. and {Roth}, Miguel and {Krisciunas}, Kevin and {Morrell}, Nidia and {Green}, E.~M. and {Persson}, S.~E. and et al.},
        title = "{An asymptotic-giant-branch star in the progenitor system of a type Ia supernova}",
      journal = {\nat},
         year = 2003,
        month = aug,
       volume = {424},
       number = {6949},
        pages = {651-654},
          doi = {10.1038/nature01854},
archivePrefix = {arXiv},
       eprint = {astro-ph/0306270},
 primaryClass = {astro-ph},
       adsurl = {https://ui.adsabs.harvard.edu/abs/2003Natur.424..651H}
}

@ARTICLE{Silverman13,
       author = {{Silverman}, Jeffrey M. and {Nugent}, Peter E. and {Gal-Yam}, Avishay and {Sullivan}, Mark and {Howell}, D. Andrew and {Filippenko}, Alexei V. and {Arcavi}, Iair and {Ben-Ami}, Sagi and {Bloom}, Joshua S. and {Cenko}, S. Bradley and et al.},
        title = "{Type Ia Supernovae Strongly Interacting with Their Circumstellar Medium}",
      journal = {\apjs},
         year = 2013,
        month = jul,
       volume = {207},
       number = {1},
          eid = {3},
        pages = {3},
          doi = {10.1088/0067-0049/207/1/3},
archivePrefix = {arXiv},
       eprint = {1304.0763},
 primaryClass = {astro-ph.CO},
       adsurl = {https://ui.adsabs.harvard.edu/abs/2013ApJS..207....3S}
}

@ARTICLE{Kirshner73,
       author = {{Kirshner}, R.~P. and {Willner}, S.~P. and {Becklin}, E.~E. and {Neugebauer}, G. and {Oke}, J.~B.},
        title = "{Spectrophotometry of the Supernova in NGC 5253 from 0.33 to 2.2 Microns}",
      journal = {\apjl},
         year = 1973,
        month = mar,
       volume = {180},
        pages = {L97},
          doi = {10.1086/181161},
       adsurl = {https://ui.adsabs.harvard.edu/abs/1973ApJ...180L..97K}
}

@ARTICLE{Wheeler98,
       author = {{Wheeler}, J. Craig and {H{\"o}flich}, Peter and {Harkness}, Robert P. and {Spyromilio}, Jason},
        title = "{Explosion Diagnostics of Type IA Supernovae from Early Infrared Spectra}",
      journal = {\apj},
         year = 1998,
        month = mar,
       volume = {496},
       number = {2},
        pages = {908-914},
          doi = {10.1086/305427},
       adsurl = {https://ui.adsabs.harvard.edu/abs/1998ApJ...496..908W}
}

@ARTICLE{Hsiao13,
       author = {{Hsiao}, E.~Y. and {Marion}, G.~H. and {Phillips}, M.~M. and {Burns}, C.~R. and {Winge}, C. and {Morrell}, N. and {Contreras}, C. and {Freedman}, W.~L. and {Kromer}, M. and {Gall}, E.~E.~E. and et al.},
        title = "{The Earliest Near-infrared Time-series Spectroscopy of a Type Ia Supernova}",
      journal = {\apj},
         year = 2013,
        month = apr,
       volume = {766},
       number = {2},
          eid = {72},
        pages = {72},
          doi = {10.1088/0004-637X/766/2/72},
archivePrefix = {arXiv},
       eprint = {1301.6287},
 primaryClass = {astro-ph.CO},
       adsurl = {https://ui.adsabs.harvard.edu/abs/2013ApJ...766...72H}
}

@ARTICLE{Sim14,
       author = {{Sim}, Chae Kyung and {Le}, Huynh Anh Nguyen and {Pak}, Soojong and {Lee}, Hye-In and {Kang}, Wonseok and {Chun}, Moo-Young and {Jeong}, Ueejeong and {Yuk}, In-Soo and {Kim}, Kang-Min and {Park}, Chan and et al.},
        title = "{Comprehensive data reduction package for the Immersion GRating INfrared Spectrograph: IGRINS}",
      journal = {Advances in Space Research},
         year = 2014,
        month = jun,
       volume = {53},
       number = {11},
        pages = {1647-1656},
          doi = {10.1016/j.asr.2014.02.024},
       adsurl = {https://ui.adsabs.harvard.edu/abs/2014AdSpR..53.1647S}
}

@ARTICLE{Sawczynec25,
       author = {{Sawczynec}, Erica and {Kaplan}, Kyle F. and {Mace}, Gregory N. and {Lee}, Jae-Joon and {Jaffe}, Daniel T. and {Park}, Chan and {Yuk}, In-Soo and {Chun}, Moo-Young and {Pak}, Soojong and {Hwang}, Narae and et al.},
        title = "{10 Years of Archival High-resolution NIR Spectra: The Raw and Reduced IGRINS Spectral Archive (RRISA)}",
      journal = {\pasp},
         year = 2025,
        month = mar,
       volume = {137},
       number = {3},
          eid = {034505},
        pages = {034505},
          doi = {10.1088/1538-3873/adba89},
archivePrefix = {arXiv},
       eprint = {2503.05867},
 primaryClass = {astro-ph.IM},
       adsurl = {https://ui.adsabs.harvard.edu/abs/2025PASP..137c4505S}
}

@MISC{Kaplan24,
       author = {{Kaplan}, Kyle and {Lee}, Jae-Joon and {Sawczynec}, Erica and {Kim}, Hyun-Jeong},
        title = "{igrins/plp}",
 howpublished = {Zenodo},
         year = 2024,
        month = apr,
      version = {3.0.0},
          doi = {10.5281/zenodo.11080095},
          eid = {10.5281/zenodo.11080095},
    publisher = {Zenodo},
       adsurl = {https://ui.adsabs.harvard.edu/abs/2024zndo..11080095K}
}

@ARTICLE{Cushing04,
       author = {{Cushing}, Michael C. and {Vacca}, William D. and {Rayner}, John T.},
        title = "{Spextool: A Spectral Extraction Package for SpeX, a 0.8-5.5 Micron Cross-Dispersed Spectrograph}",
      journal = {\pasp},
         year = 2004,
        month = apr,
       volume = {116},
       number = {818},
        pages = {362-376},
          doi = {10.1086/382907},
       adsurl = {https://ui.adsabs.harvard.edu/abs/2004PASP..116..362C}
}

@ARTICLE{Hsiao07,
       author = {{Hsiao}, E.~Y. and {Conley}, A. and {Howell}, D.~A. and {Sullivan}, M. and {Pritchet}, C.~J. and {Carlberg}, R.~G. and {Nugent}, P.~E. and {Phillips}, M.~M.},
        title = "{K-Corrections and Spectral Templates of Type Ia Supernovae}",
      journal = {\apj},
         year = 2007,
        month = jul,
       volume = {663},
       number = {2},
        pages = {1187-1200},
          doi = {10.1086/518232},
archivePrefix = {arXiv},
       eprint = {astro-ph/0703529},
 primaryClass = {astro-ph},
       adsurl = {https://ui.adsabs.harvard.edu/abs/2007ApJ...663.1187H}
}

@ARTICLE{Lu23,
       author = {{Lu}, Jing and {Hsiao}, Eric Y. and {Phillips}, Mark M. and {Burns}, Christopher R. and {Ashall}, Chris and {Morrell}, Nidia and {Ng}, Lawrence and {Kumar}, Sahana and {Shahbandeh}, Melissa and {Hoeflich}, Peter and et al.},
        title = "{Carnegie Supernova Project. II. Near-infrared Spectral Diversity and Template of Type Ia Supernovae}",
      journal = {\apj},
         year = 2023,
        month = may,
       volume = {948},
       number = {1},
          eid = {27},
        pages = {27},
          doi = {10.3847/1538-4357/acc100},
archivePrefix = {arXiv},
       eprint = {2211.05998},
 primaryClass = {astro-ph.HE},
       adsurl = {https://ui.adsabs.harvard.edu/abs/2023ApJ...948...27L}
}

@ARTICLE{Shen13,
       author = {{Shen}, Ken J. and {Guillochon}, James and {Foley}, Ryan J.},
        title = "{Circumstellar Absorption in Double Detonation Type Ia Supernovae}",
      journal = {\apjl},
         year = 2013,
        month = jun,
       volume = {770},
       number = {2},
          eid = {L35},
        pages = {L35},
          doi = {10.1088/2041-8205/770/2/L35},
archivePrefix = {arXiv},
       eprint = {1302.2916},
 primaryClass = {astro-ph.SR},
       adsurl = {https://ui.adsabs.harvard.edu/abs/2013ApJ...770L..35S}
}

@ARTICLE{Marion09,
       author = {{Marion}, G.~H. and {H{\"o}flich}, P. and {Gerardy}, C.~L. and {Vacca}, W.~D. and {Wheeler}, J.~C. and {Robinson}, E.~L.},
        title = "{A Catalog of Near-Infrared Spectra from Type Ia Supernovae}",
      journal = {\aj},
         year = 2009,
        month = sep,
       volume = {138},
       number = {3},
        pages = {727-757},
          doi = {10.1088/0004-6256/138/3/727},
archivePrefix = {arXiv},
       eprint = {0906.4085},
 primaryClass = {astro-ph.CO},
       adsurl = {https://ui.adsabs.harvard.edu/abs/2009AJ....138..727M}
}

@ARTICLE{Sternberg14,
       author = {{Sternberg}, A. and {Gal-Yam}, A. and {Simon}, J.~D. and {Patat}, F. and {Hillebrandt}, W. and {Phillips}, M.~M. and {Foley}, R.~J. and {Thompson}, I. and {Morrell}, N. and {Chomiuk}, L. and {Soderberg}, A.~M. and {Yong}, D. and {Kraus}, A.~L. and {Herczeg}, G.~J. and {Hsiao}, E.~Y. and {Raskutti}, S. and {Cohen}, J.~G. and {Mazzali}, P.~A. and {Nomoto}, K.},
        title = "{Multi-epoch high-spectral-resolution observations of neutral sodium in 14 Type Ia supernovae}",
      journal = {\mnras},
         year = 2014,
        month = sep,
       volume = {443},
       number = {2},
        pages = {1849-1860},
          doi = {10.1093/mnras/stu1202},
archivePrefix = {arXiv},
       eprint = {1311.3645},
 primaryClass = {astro-ph.HE},
       adsurl = {https://ui.adsabs.harvard.edu/abs/2014MNRAS.443.1849S}
}

@ARTICLE{Maeda16,
       author = {{Maeda}, K. and {Tajitsu}, A. and {Kawabata}, K.~S. and {Foley}, R.~J. and {Honda}, S. and {Moritani}, Y. and {Tanaka}, M. and {Hashimoto}, O. and {Ishigaki}, M. and {Simon}, J.~D. and {Phillips}, M.~M. and {Yamanaka}, M. and {Nogami}, D. and {Arai}, A. and {Aoki}, W. and {Nomoto}, K. and {Milisavljevic}, D. and {Mazzali}, P.~A. and {Soderberg}, A.~M. and {Schramm}, M. and {Sato}, B. and {Harakawa}, H. and {Morrell}, N. and {Arimoto}, N.},
        title = "{Sodium Absorption Systems toward SN Ia 2014J Originate on Interstellar Scales}",
      journal = {\apj},
         year = 2016,
        month = jan,
       volume = {816},
       number = {2},
          eid = {57},
        pages = {57},
          doi = {10.3847/0004-637X/816/2/57},
archivePrefix = {arXiv},
       eprint = {1511.05668},
 primaryClass = {astro-ph.SR},
       adsurl = {https://ui.adsabs.harvard.edu/abs/2016ApJ...816...57M}
}

@ARTICLE{Ferretti16,
       author = {{Ferretti}, R. and {Amanullah}, R. and {Goobar}, A. and {Johansson}, J. and {Vreeswijk}, P.~M. and {Butler}, R.~P. and {Cao}, Y. and {Cenko}, S.~B. and {Doran}, G. and {Filippenko}, A.~V. and {Freeland}, E. and {Hosseinzadeh}, G. and {Howell}, D.~A. and {Lundqvist}, P. and {Mattila}, S. and {Nordin}, J. and {Nugent}, P.~E. and {Petrushevska}, T. and {Valenti}, S. and {Vogt}, S. and {Wozniak}, P.},
        title = "{Time-varying sodium absorption in the Type Ia supernova 2013gh}",
      journal = {\aap},
         year = 2016,
        month = jul,
       volume = {592},
          eid = {A40},
        pages = {A40},
          doi = {10.1051/0004-6361/201628351},
archivePrefix = {arXiv},
       eprint = {1605.01738},
 primaryClass = {astro-ph.SR},
       adsurl = {https://ui.adsabs.harvard.edu/abs/2016A&A...592A..40F}
}

@ARTICLE{Fulton15,
       author = {{Fulton}, Benjamin J. and {Weiss}, Lauren M. and {Sinukoff}, Evan and {Isaacson}, Howard and {Howard}, Andrew W. and {Marcy}, Geoffrey W. and {Henry}, Gregory W. and {Holden}, Bradford P. and {Kibrick}, Robert I.},
        title = "{Three Super-Earths Orbiting HD 7924}",
      journal = {\apj},
         year = 2015,
        month = jun,
       volume = {805},
       number = {2},
          eid = {175},
        pages = {175},
          doi = {10.1088/0004-637X/805/2/175},
archivePrefix = {arXiv},
       eprint = {1504.06629},
 primaryClass = {astro-ph.EP},
       adsurl = {https://ui.adsabs.harvard.edu/abs/2015ApJ...805..175F}
}

@article{astropy:2013,
  author  = {{Astropy Collaboration} and {Robitaille}, T.~P. and {Tollerud}, E.~J. and others},
  title   = {{Astropy: A community Python package for astronomy}},
  journal = {A\&A}, year = 2013, volume = 558, eid = {A33}, pages = {A33},
  doi = {10.1051/0004-6361/201322068},
  adsurl  = {https://ui.adsabs.harvard.edu/abs/2013A&A...558A..33A}
}

@article{astropy:2018,
  author  = {{Astropy Collaboration} and {Price-Whelan}, A.~M. and {Sip{\H o}cz}, B.~M. and others},
  title   = {{The Astropy Project: Building an Open-science Project and Status of the v2.0 Core Package}},
  journal = {AJ}, year = 2018, volume = 156, eid = {123}, pages = {123},
  doi = {10.3847/1538-3881/aabc4f},
  adsurl  = {https://ui.adsabs.harvard.edu/abs/2018AJ....156..123A}
}

@article{astropy:2022,
  author  = {{Astropy Collaboration} and {Price-Whelan}, A.~M. and {Lim}, P.~L. and others},
  title   = {{The Astropy Project: Sustaining and Growing a Community-oriented Open-source Project and the Latest Major Release (v5.0) of the Core Package}},
  journal = {ApJ}, year = 2022, volume = 935, eid = {167}, pages = {167},
  doi = {10.3847/1538-4357/ac7c74},
  adsurl  = {https://ui.adsabs.harvard.edu/abs/2022ApJ...935..167A}
}

@article{Hunter:2007,
  author  = {{Hunter}, J.~D.},
  title   = {{Matplotlib: A 2D Graphics Environment}},
  journal = {Computing in Science and Engineering}, year = 2007,
  volume = 9, number = 3, pages = {90--95},
  doi = {10.1109/MCSE.2007.55},
  adsurl  = {https://ui.adsabs.harvard.edu/abs/2007CSE.....9...90H}
}

@article{harris2020array,
  author  = {{Harris}, C.~R. and {Millman}, K.~J. and {van der Walt}, S.~J. and others},
  title   = {{Array programming with NumPy}},
  journal = {Nature}, year = 2020, volume = 585, number = 7825, pages = {357--362},
  doi = {10.1038/s41586-020-2649-2},
  adsurl  = {https://ui.adsabs.harvard.edu/abs/2020Natur.585..357H}
}

@article{2020SciPy-NMeth,
  author  = {{Virtanen}, P. and {Gommers}, R. and {Oliphant}, T.~E. and others},
  title   = {{SciPy 1.0: Fundamental Algorithms for Scientific Computing in Python}},
  journal = {Nature Methods}, year = 2020, volume = 17, pages = {261--272},
  doi = {10.1038/s41592-019-0686-2},
  adsurl  = {https://ui.adsabs.harvard.edu/abs/2020NatMe..17..261V}
}

@software{reback2020pandas,
  author    = {{The pandas development team}},
  title     = {{pandas-dev/pandas: Pandas}},
  year      = 2020, publisher = {Zenodo},
  doi = {10.5281/zenodo.3509134},
  adsurl    = {https://ui.adsabs.harvard.edu/abs/2022zndo...3509134T}
}

@misc{pya,
  author  = {{Czesla}, S. and {Schr{\"o}ter}, S. and {Schneider}, C.~P. and {Huber}, K.~F. and {Pfeifer}, F. and {Andreasen}, D.~T. and {Zechmeister}, M.},
  title   = {{PyA: Python astronomy-related packages}}, year = 2019, month = jun,
  eid = {ascl:1906.010}, pages = {ascl:1906.010},
  archivePrefix = {ascl}, eprint = {1906.010},
  adsurl  = {https://ui.adsabs.harvard.edu/abs/2019ascl.soft06010C}
}

@inproceedings{mckinney-proc-scipy-2010,
  author    = {{McKinney}, Wes},
  title     = {{Data Structures for Statistical Computing in Python}},
  booktitle = {Proceedings of the 9th Python in Science Conference},
  year      = 2010, pages = {56--61},
  doi = {10.25080/Majora-92bf1922-00a}
}

@ARTICLE{Ferretti17a,
       author = {{Ferretti}, R. and {Amanullah}, R. and {Goobar}, A. and {Petrushevska}, T. and {Borthakur}, S. and {Bulla}, M. and {Fox}, O. and {Freeland}, E. and {Fremling}, C. and {Hangard}, L. and {Hayes}, M.},
        title = "{Probing gas and dust in the tidal tail of NGC 5221 with the type Ia supernova iPTF16abc}",
      journal = {\aap},
         year = 2017,
        month = oct,
       volume = {606},
          eid = {A111},
        pages = {A111},
          doi = {10.1051/0004-6361/201731409},
archivePrefix = {arXiv},
       eprint = {1708.07133},
 primaryClass = {astro-ph.GA},
       adsurl = {https://ui.adsabs.harvard.edu/abs/2017A&A...606A.111F}
}

@ARTICLE{Ferretti17b,
       author = {{Ferretti}, Raphael and {Amanullah}, Rahman and {Bulla}, Mattia and {Goobar}, Ariel and {Johansson}, Joel and {Lundqvist}, Peter},
        title = "{No Evidence of Circumstellar Gas Surrounding Type Ia Supernova SN 2017cbv}",
      journal = {\apjl},
         year = 2017,
        month = dec,
       volume = {851},
       number = {2},
          eid = {L43},
        pages = {L43},
          doi = {10.3847/2041-8213/aa9e49},
archivePrefix = {arXiv},
       eprint = {1708.05394},
 primaryClass = {astro-ph.SR},
       adsurl = {https://ui.adsabs.harvard.edu/abs/2017ApJ...851L..43F}
}

@ARTICLE{Gonzalez-Gaitan24,
       author = {{Gonz{\'a}lez-Gait{\'a}n}, Santiago and {Guti{\'e}rrez}, Claudia P. and {Anderson}, Joseph P. and {Morales-Garoffolo}, Antonia and {Galbany}, Lluis and {Goswami}, Sabyasachi and {Mour{\~a}o}, Ana M. and {Mattila}, Seppo and {Sullivan}, Mark},
        title = "{Narrow absorption lines from intervening material in supernovae. I. Measurements and temporal evolution}",
      journal = {\aap},
         year = 2024,
        month = jul,
       volume = {687},
          eid = {A108},
        pages = {A108},
          doi = {10.1051/0004-6361/202348818},
archivePrefix = {arXiv},
       eprint = {2403.11677},
 primaryClass = {astro-ph.IM},
       adsurl = {https://ui.adsabs.harvard.edu/abs/2024A&A...687A.108G}
}

@ARTICLE{Sollerman05,
       author = {{Sollerman}, J. and {Cox}, N. and {Mattila}, S. and {Ehrenfreund}, P. and {Kaper}, L. and {Leibundgut}, B. and {Lundqvist}, P.},
        title = "{Diffuse Interstellar Bands in <ASTROBJ>NGC 1448</ASTROBJ>}",
      journal = {\aap},
         year = 2005,
        month = jan,
       volume = {429},
        pages = {559-567},
          doi = {10.1051/0004-6361:20041465},
archivePrefix = {arXiv},
       eprint = {astro-ph/0409340},
 primaryClass = {astro-ph},
       adsurl = {https://ui.adsabs.harvard.edu/abs/2005A&A...429..559S}
}

@ARTICLE{Riess98,
       author = {{Riess}, Adam G. and {Filippenko}, Alexei V. and {Challis}, Peter and {Clocchiatti}, Alejandro and {Diercks}, Alan and {Garnavich}, Peter M. and {Gilliland}, Ron L. and {Hogan}, Craig J. and {Jha}, Saurabh and {Kirshner}, Robert P. and {Leibundgut}, B. and {Phillips}, M.~M. and {Reiss}, David and {Schmidt}, Brian P. and {Schommer}, Robert A. and {Smith}, R. Chris and {Spyromilio}, J. and {Stubbs}, Christopher and {Suntzeff}, Nicholas B. and {Tonry}, John},
        title = "{Observational Evidence from Supernovae for an Accelerating Universe and a Cosmological Constant}",
      journal = {\aj},
         year = 1998,
        month = sep,
       volume = {116},
       number = {3},
        pages = {1009-1038},
          doi = {10.1086/300499},
archivePrefix = {arXiv},
       eprint = {astro-ph/9805201},
 primaryClass = {astro-ph},
       adsurl = {https://ui.adsabs.harvard.edu/abs/1998AJ....116.1009R}
}

@ARTICLE{Munari_Zwitter97,
       author = {{Munari}, U. and {Zwitter}, T.},
        title = "{Equivalent width of NA I and K I lines and reddening.}",
      journal = {\aap},
         year = 1997,
        month = feb,
       volume = {318},
        pages = {269-274},
       adsurl = {https://ui.adsabs.harvard.edu/abs/1997A&A...318..269M}
}

@ARTICLE{Kwok26,
       author = {{Kwok}, Lindsey A. and {Blondin}, St{\'e}phane and {Miller}, Adam A. and {Jha}, Saurabh W. and {Hoogendam}, Willem B. and {Pfeffer}, Cameron M. and {Abate}, Eyouel Z. and {Andrews}, Jennifer E. and {Andrews}, Moira and {Ashall}, Chris and {Auchettl}, Katie and {Bostroem}, K. Azalee and {Brink}, Thomas G. and {Callan}, Fionntan P. and {Christy}, Collin T. and {Farah}, Joseph R. and {Filippenko}, Alexei V. and {Fl{\"o}rs}, Andreas and {Foley}, Ryan J. and {Gendreau-Distler}, Eli and {Graur}, Or and {Hinkle}, Jason T. and {Howell}, D. Andrew and {Jones}, David O. and {Kageyama}, Rinon and {Kawabata}, Miho and {Koelln}, Cristine and {Larison}, Conor and {Liu}, Chang and {Maeda}, Keiichi and {Maguire}, Kate and {McCully}, Curtis and {Medler}, Kyle and {Meza-Retamal}, Nicolas E. and {Mina}, Ann and {Pakmor}, R{\"u}diger and {Patlak}, Riley and {Pearson}, Jeniveve and {Ravi}, Aravind P. and {Rehemtulla}, Nabeel and {Rest}, Armin and {Sand}, David J. and {Sears}, Huei and {Shappee}, Benjamin J. and {Shrestha}, Manisha and {Singh}, Mridweeka and {Szalai}, Tam{\'a}s and {Taguchi}, Kenta and {Temim}, Tea and {Terwel}, Jacco H. and {Valenti}, Stefano and {Vink{\'o}}, J{\'o}zsef and {Wheeler}, J. Craig and {Wynn}, Kathryn and {Yang}, Yi and {Zheng}, WeiKang},
        title = "{JWST Spectroscopy of Type Ia Supernova 2025rbs from Maximum Light to the Nebular Phase}",
      journal = {arXiv e-prints},
         year = 2026,
        month = aug,
          eid = {arXiv:2608.10451},
        pages = {arXiv:2608.10451},
archivePrefix = {arXiv},
       eprint = {2608.10451},
 primaryClass = {astro-ph.HE},
       adsurl = {https://ui.adsabs.harvard.edu/abs/2026arXiv260810451K}
}

@ARTICLE{Gonzalez-Gaitan25,
       author = {{Gonz{\'a}lez-Gait{\'a}n}, Santiago and {Guti{\'e}rrez}, Claudia P. and {Martins}, Gon{\c{c}}alo and {M{\"u}ller-Bravo}, Tom{\'a}s E. and {Duarte}, Jo{\~a}o and {Anderson}, Joseph P. and {Galbany}, Lluis and {Sullivan}, Mark and {Rino-Silvestre}, Jo{\~a}o and {Caixach}, Mariona and {Morales-Garoffolo}, Antonia and {Goswami}, Sabyasachi and {Mour{\~a}o}, Ana M. and {Mattila}, Seppo},
        title = "{Narrow absorption lines from intervening material in supernovae: II. Galaxy properties}",
      journal = {\aap},
         year = 2025,
        month = aug,
       volume = {700},
          eid = {A119},
        pages = {A119},
          doi = {10.1051/0004-6361/202554355},
archivePrefix = {arXiv},
       eprint = {2503.07233},
 primaryClass = {astro-ph.GA},
       adsurl = {https://ui.adsabs.harvard.edu/abs/2025A&A...700A.119G}
}

@ARTICLE{Thilker07,
       author = {{Thilker}, David A. and {Boissier}, Samuel and {Bianchi}, Luciana and {Calzetti}, Daniela and {Boselli}, Alessandro and {Dale}, Daniel A. and {Seibert}, Mark and {Braun}, Robert and {Burgarella}, Denis and {Gil de Paz}, Armando and {Helou}, George and {Walter}, Fabian and {Kennicutt}, Jr., R.~C. and {Madore}, Barry F. and {Martin}, D. Christopher and {Barlow}, Tom A. and {Forster}, Karl and {Friedman}, Peter G. and {Morrissey}, Patrick and {Neff}, Susan G. and {Schiminovich}, David and {Small}, Todd and {Wyder}, Ted K. and {Donas}, Jos{\'e} and {Heckman}, Timothy M. and {Lee}, Young-Wook and {Milliard}, Bruno and {Rich}, R. Michael and {Szalay}, Alex S. and {Welsh}, Barry Y. and {Yi}, Sukyoung K.},
        title = "{Ultraviolet and Infrared Diagnostics of Star Formation and Dust in NGC 7331}",
      journal = {\apjs},
         year = 2007,
        month = dec,
       volume = {173},
       number = {2},
        pages = {572-596},
          doi = {10.1086/516646},
       adsurl = {https://ui.adsabs.harvard.edu/abs/2007ApJS..173..572T}
}

@ARTICLE{Telesco82,
       author = {{Telesco}, C.~M. and {Gatley}, I. and {Stewart}, J.~M.},
        title = "{The distribution of infrared obscuration in NGC 7331 - Evidence for a massive molecular ring}",
      journal = {\apjl},
         year = 1982,
        month = dec,
       volume = {263},
        pages = {L13-L17},
          doi = {10.1086/183914},
       adsurl = {https://ui.adsabs.harvard.edu/abs/1982ApJ...263L..13T}
}

@ARTICLE{Smith04,
       author = {{Smith}, J.~D.~T. and {Dale}, D.~A. and {Armus}, L. and {Draine}, B.~T. and {Hollenbach}, D.~J. and {Roussel}, H. and {Helou}, G. and {Kennicutt}, Jr., R.~C. and {Li}, A. and {Bendo}, G.~J. and {Calzetti}, D. and {Engelbracht}, C.~W. and {Gordon}, K.~D. and {Jarrett}, T.~H. and {Kewley}, L. and {Leitherer}, C. and {Malhotra}, S. and {Meyer}, M.~J. and {Murphy}, E.~J. and {Regan}, M.~W. and {Rieke}, G.~H. and {Rieke}, M.~J. and {Thornley}, M.~D. and {Walter}, F. and {Wolfire}, M.~G.},
        title = "{Mid-Infrared IRS Spectroscopy of NGC 7331: A First Look at the Spitzer Infrared Nearby Galaxies Survey (SINGS) Legacy}",
      journal = {\apjs},
         year = 2004,
        month = sep,
       volume = {154},
       number = {1},
        pages = {199-203},
          doi = {10.1086/423133},
archivePrefix = {arXiv},
       eprint = {astro-ph/0406332},
 primaryClass = {astro-ph},
       adsurl = {https://ui.adsabs.harvard.edu/abs/2004ApJS..154..199S}
}

@ARTICLE{Schlafly_Finkbeiner11,
       author = {{Schlafly}, Edward F. and {Finkbeiner}, Douglas P.},
        title = "{Measuring Reddening with Sloan Digital Sky Survey Stellar Spectra and Recalibrating SFD}",
      journal = {\apj},
         year = 2011,
        month = aug,
       volume = {737},
       number = {2},
          eid = {103},
        pages = {103},
          doi = {10.1088/0004-637X/737/2/103},
archivePrefix = {arXiv},
       eprint = {1012.4804},
 primaryClass = {astro-ph.GA},
       adsurl = {https://ui.adsabs.harvard.edu/abs/2011ApJ...737..103S}
}
\bibliographystyle{aasjournalv7}

\end{CJK*}
\end{document}